\documentclass[10pt,preprint]{emulateapj}

\newcommand*{\teff}{$T_{\rm eff}$}
\newcommand*{\logg}{$\log~g$}
\newcommand*{\feh}{[Fe/H]}
\newcommand*{\afe}{[$\alpha$/Fe]}
\newcommand*{\cfe}{[C/Fe]}

\newcommand*{\alp}{$\alpha$}

\usepackage{amsmath}
\usepackage{amssymb}
\usepackage{amsfonts}
\usepackage{longtable}
\usepackage{apjfonts}
\usepackage{graphics}
\usepackage{natbib}

\shorttitle{CARBON-ENHANCED METAL-POOR STARS IN SDSS/SEGUE. I}
\shortauthors{Lee et al.}
\slugcomment{Accepted on August 20, 2013}

\begin{document}

\title{Carbon-enhanced Metal-poor Stars in SDSS/SEGUE. I. \\
Carbon Abundance Estimation and Frequency of CEMP Stars}

\author{Young Sun Lee\altaffilmark{1}, Timothy C. Beers\altaffilmark{2,3},
Thomas Masseron\altaffilmark{4}, Bertrand Plez\altaffilmark{5}, 
Constance M. Rockosi\altaffilmark{6}, \\ Jennifer Sobeck\altaffilmark{7,8},
Brian Yanny\altaffilmark{9}, Sara Lucatello\altaffilmark{10}, 
Thirupathi Sivarani\altaffilmark{11}, \\ Vinicius M. Placco\altaffilmark{2,12}, 
and Daniela Carollo\altaffilmark{13} \\
\scriptsize{
\textup{
\affil{1}{$^1$Department of Astronomy, New Mexico State University,
                 Las Cruces, NM, 88003, USA; yslee@nmsu.edu} \\
\affil{2}{$^2$National Optical Astronomy Observatory, Tucson, AZ 85719, USA} \\
\affil{3}{$^3$Joint Institute for Nuclear Astrophysics (JINA), Michigan State 
                 University, East Lansing, MI 48824, USA} \\
\affil{4}{$^4$Institut d'Astronomie et d'Astrophysique, Universit\'e Libre de
                 Bruxelles, CP 226, Boulevard du Triomphe, B-1050 Bruxelles, Belgium} \\
\affil{5}{$^5$Laboratoire Univers et Particules de Montpellier, Universit\'e
                 Montpellier 2, CNRS, F-34095 Montpellier, France} \\
\affil{6}{$^6$UCO/Lick Observatory, Department of Astronomy and Astrophysics,
                 University of California, Santa Cruz, CA 95064, USA} \\
\affil{7}{$^7$Laboratoire Lagrange (UMR7293), Universit\'e de Nice Sophia 
                 Antipolis, CNRS, Observatoire de la C\^ote d'Azur, BP 4229, 
                 F-06304 Nice Cedex 04, France} \\
\affil{8}{$^8$JINA--Joint Institute for Nuclear Astrophysics and the Department 
                 of Astronomy and Astrophysics, University of Chicago, 5640 South 
                 Ellis Avenue, Chicago, IL 60637, USA} \\
\affil{9}{$^9$Fermi National Accelerator Laboratory, Batavia, IL 60510, USA} \\
\affil{10}{$^{10}$INAF--Osservatorio Astronomico di Padova, vicolo 
                 dell'Osservatorio 5, I-35122 Padova, Italy} \\
\affil{11}{$^{11}$Indian Institute of Astrophysics, 2nd block Koramangala, 
                  Bangalore-560034, India} \\
\affil{12}{$^{12}$Departamento de Astronomia, Instituto de Astronomia, 
                  Geof\'isica~e Ci\^ encias Atmosf\'ericas, Universidade de S\~ ao 
                  Paulo, S\~ ao Paulo, SP 05508-090, Brazil} \\
\affil{13}{$^{13}$Department of Physics and Astronomy, Astronomy, Astrophysics and 
                  Astrophotonic Research Center, Macquarie University, North 
                  Ryde, NSW, 2019, Australia}}}}

\begin{abstract}
We describe a method for the determination of stellar [C/Fe] abundance ratios
using low-resolution ($R = 2000$) stellar spectra from the Sloan Digital
Sky Survey (SDSS) and its Galactic sub-survey, the Sloan Extension for
Galactic Understanding and Exploration (SEGUE). By means of a
star-by-star comparison with a set of SDSS/SEGUE spectra with available
estimates of \cfe\ based on published high-resolution analyses, we
demonstrate that we can measure [C/Fe] from SDSS/SEGUE spectra with S/N
$\geq 15$ \AA$^{-1}$ to a precision better than 0.35 dex for stars with
atmospheric parameters in the range \teff~= [4400, 6700] K, \logg~=
[1.0, 5.0], \feh~=~[$-$4.0, $+$0.5], and \cfe~=~[$-0.25$, $+$3.5]. Using 
the measured carbon-to-iron abundance ratios obtained by this
technique, we derive the frequency of carbon-enhanced stars (\cfe\ $\geq
+0.7$) as a function of [Fe/H], for both the SDSS/SEGUE stars and other
samples from the literature. We find that the differential frequency
slowly rises from almost zero to about 14\% at [Fe/H] $\sim$ --2.4,
followed by a sudden increase, by about a factor of three, to 39\% from
[Fe/H] $\sim$ --2.4 to [Fe/H] $\sim$ --3.7. Although the number of stars
known with [Fe/H] $< -4.0$ remains small, the frequency of
carbon-enhanced metal-poor (CEMP) stars below this value is around 75\%. We 
also examine how the cumulative frequency of CEMP stars varies across
different luminosity classes. The giant sample exhibits a cumulative
CEMP frequency of 32\% for \feh\ $\leq -2.5$, 31\% for \feh\ $\leq
-3.0$, and 33\% for \feh\ $\leq -3.5$; a roughly constant value. 
For the main-sequence turnoff stars, we obtain a lower cumulative CEMP
frequency, around 10\% for \feh\ $\leq -2.5$, presumably due
to the difficulty of identifying CEMP stars among warmer turnoff stars
with weak CH $G$-bands. The dwarf population displays a large
change in the cumulative frequency for CEMP stars below [Fe/H] = --2.5,
jumping from 15\% for \feh\ $\leq -2.5$ to about 75\% for \feh\ $\leq
-3.0$. When we impose a restriction with respect to distance from the Galactic
mid-plane ($|Z| < 5$ kpc), the frequency of the CEMP giants does not
increase at low metallicity (\feh\ $<-2.5$), but rather, decreases, due
to the dilution of C-rich material in stars that have undergone mixing
with CNO-processed material from their interiors. The frequency of CEMP
stars near the main-sequence turnoff, which are not expected to have
experienced mixing, increases for \feh\ $\leq -3.0$. The general rise in
the global CEMP frequency at low metallicity is likely due to the
transition from the inner-halo to the outer-halo stellar populations
with declining metallicity and increasing distance from the plane.
\end{abstract}

\keywords{methods: data analysis -- stars: abundances, fundamental
parameters -- surveys -- techniques: imaging spectroscopy}

\section{Introduction}

The chemical abundance ratios of very metal-poor (VMP;
[Fe/H]\footnote[14]{[Fe/H]=log$_{10}$($N$(Fe)/$N$(H))$_\star$ --
log$_{10}$($N$(Fe)/$N$(H))$_\odot$}$\leq -2.0$) stars in the Milky Way
provide vital clues to the early chemical evolution and initial mass
function (IMF) of their progenitors, which are likely to have been among
the first generations of stars formed in the universe.

Given this importance, there have been increasingly ambitious efforts
carried out to identify metal-poor candidates with large-scale surveys
of the stellar populations of the Milky Way. The early HK survey (Beers
et al. 1985, 1992) and the Hamburg/ESO Survey (HES; Wisotzki et al.
1996; Christlieb 2003; Christlieb et al. 2001, 2008) collectively identified
several thousand VMP stars (Beers \& Christlieb 2005). In recent years,
this number has been dramatically increased by the Sloan Digital Sky
Survey (SDSS; Fukugita et al. 1996; Gunn et al. 1998, 2006; York et al.
2000; Stoughton et al. 2002; Abazajian et al. 2003, 2004, 2005, 2009;
Pier et al. 2003; Adelman-McCarthy et al. 2006, 2007, 2008; Aihara et
al. 2011; Ahn et al. 2012) and the Sloan Extension for Galactic
Understanding and Exploration (SEGUE-1; Yanny et al. 2009) and SEGUE-2
(C. Rockosi et al., in preparation), to many tens of thousands of VMP
stars. Ongoing surveys, such as Large sky Area Multi-Object fiber 
Spectroscopic Telescope (LAMOST; Cui et al. 2012; Luo et al.
2012), hold the promise of enlarging this sample to several hundred
thousand VMP stars.

Detailed chemical-abundance analyses by a number of groups, based on
high-resolution spectroscopic follow-up, have revealed that, while most
VMP stars exhibit similar abundance patterns, there are numerous
examples of objects with peculiar chemical patterns, such as strong
enrichments or deficiencies of light elements such as C, N, O, Na, Mg,
Si, etc. (e.g., McWilliam et al. 1995; Ryan et al. 1996; Norris et al.
2001, 2013; Johnson 2002; Cayrel et al. 2004; Aoki et al. 2008, 2013;
Cohen et al. 2008; Lai et al. 2008). Frebel \& Norris (2012) provide a
useful summary of all but the most recent work. Among the chemically
peculiar stars with \feh\ $\leq -2.0$, objects with enhanced carbon
abundance are the most common variety. 

The carbon-enhancement phenomenon was recognized over a half century ago
for stars with [Fe/H] $\sim$ --2.0 to [Fe/H] $\sim -1.0$. Such objects
were called CH stars (Keenan 1942) or subgiant CH stars (Bond 1974),
because their optical spectra exhibit strong CH $G$-band absorption
features around 4300 \AA\ compared to stars with similar effective
temperatures and metallicities. Over the past quarter century,
spectroscopic follow-up of metal-poor candidates selected from the HK
and HES surveys have identified many more such stars at even lower
metallicity (e.g., Beers et al. 1985, 1992; Christlieb et al. 2001;
Christlieb 2003). These stars are referred to as carbon-enhanced
metal-poor (CEMP) stars, originally defined as having metallicity [Fe/H]
$\leq -1.0$ and carbon-to-iron ratios larger than 10 times the solar
ratio (i.e., [C/Fe]\footnote[15]{[C/Fe]=log$_{10}$($N$(C) /$N$(Fe))
$_\star$ -- log$_{10}$($N$(C)/$N$(Fe))$_\odot$} $> +1.0$; Beers
\& Christlieb 2005); definitions of CEMP stars using the criteria 
[C/Fe] $> +0.5$ and [C/Fe] $> +0.7$ have also been employed by a number
of authors (e.g., Aoki et al. 2007).

Recent studies of VMP stars discovered from various spectroscopic
surveys have confirmed that the cumulative fraction of CEMP stars
strongly increases with decreasing metallicity. Overall, the cumulative
fraction of CEMP stars rises from $\sim$ 20 \% for [Fe/H] $< -2.0$, 30\%
for [Fe/H] $< -3.0$, 40\% for [Fe/H] $< -3.5$, and 75\% for [Fe/H] $<
-4.0$, after the inclusion of a carbon-normal star with \feh\ $\sim
-5.0$ (Beers \& Christlieb 2005; Marsteller et al. 2005; Rossi et al.
2005; Frebel et al. 2006; Lucatello et al. 2006; Norris et al. 2007,
2013; Carollo et al. 2012; Spite et al. 2013; Yong et al. 2013). 

The high frequency of CEMP stars at low metallicity indicates that a
large amount of carbon (relative to iron) was produced at an early
evolutionary stage of the Milky Way. Other evidence for the large 
production of carbon at early times comes from the discovery by Cooke et al. (2011) 
of an extremely metal-poor (EMP; [Fe/H] $\sim –3.0$) damped Ly-$\alpha$ (DLA) 
system at $z = 2.3$ that exhibits enhanced a carbon abundance ratio 
([C/Fe]$ = +1.5$) and other elemental-abundance signatures similar to 
the CEMP-no class of stars. This classification (defined by Beers \& Christlieb 2005) 
describes CEMP stars with no strong enhancements of $s$-process elements. Matsuoka et
al. (2011) also reported evidence for strong carbon production in the
early universe, based on their analysis of the optical spectrum of the
most distant known radio galaxy, with $z$ = 5.19. 

The mechanisms that have been proposed to account for this large carbon
production include: (1) mass transfer of carbon-enhanced material from
the envelope of an asymptotic giant branch (AGB) star to its (presently
observed) binary companion (e.g., Herwig 2005; Sneden et al. 2008;
Masseron et al. 2010; Bisterzo et al. 2011, 2012); (2) massive, rapidly
rotating, zero-metallicity stars, which produce large amounts of carbon,
nitrogen, and oxygen due to distinctive internal burning and mixing
episodes (Meynet et al. 2006, 2010); and (3) faint supernovae associated
with the first generations of stars, which experience extensive mixing
and fallback during their explosions and eject large amounts of C and O
(Umeda \& Nomoto 2003, 2005; Tominaga et al. 2007; Ito et al. 2009,
2013). 

This early carbon production can have a profound influence on the
chemical evolution of the Galaxy and the universe. As argued by Abia et
al. (2001) and Lucatello et al. (2005), and most recently by Suda et al.
(2013), the large fraction of CEMP stars at the lowest metallicities
could indicate that the IMF in the early universe included a larger
number of intermediate- to high-mass stars than the present-day IMF.
However, Izzard et al. (2009) and Abate et al. (2013), using a binary
population model alone, could not reproduce the high fraction of CEMP
stars among the most metal-poor stars.

In order to understand the carbon-enhanced phenomenon among 
low-metallicity stars, there have been a number of efforts to estimate
carbon abundance from low- and medium-resolution stellar spectroscopic
surveys of metal-poor candidates. Generally, these efforts were limited
in terms of sample size, coverage of stellar parameter space, and
methodology for determining \cfe. For example, the early approach of
Rossi et al. (2005) made use of the strength of the Ca\,{\sc ii} K line
and CH $G$-band, along with the associated broadband $J-K$ colors, to
estimate [Fe/H] and [C/Fe] for medium-resolution (1--2 \AA) spectra of
HK-survey stars. Frebel et al. (2006) followed Rossi et al.'s
prescription to obtain \cfe\ for a subset of 234 stars among 1777
metal-poor candidates from the HES survey, deriving a frequency of
9\%$\pm$2\% for VMP giants with \cfe\ $> +1.0$. Very recently, Carollo et
al. (2012) employed a grid of synthetic spectra, covering a wide range
of parameter space, in order to match with the SDSS/SEGUE stellar
spectra around the CH $G$-band region to derive the carbon-to-iron ratios.
They applied their technique to the calibration stars (used for
spectrophotometric corrections and tests of interstellar reddening) from
SDSS/SEGUE, ending up with about 31,000 stars with derived carbon
abundances (or upper limits), the largest previous effort to measure
[C/Fe] for halo stars to date.

The SDSS/SEGUE surveys have produced an unprecedented sample of stellar
spectra for more detailed analysis. SEGUE-1 was one of three sub-surveys
comprising SDSS-II (Legacy, Supernova Survey, and SEGUE-1). The SEGUE-1
program obtained $ugriz$ imaging of some 3500 deg$^{2}$ of sky outside
of the original SDSS footprint, and roughly 240,000 low-resolution
($R=2000$) stellar spectra covering the wavelength range 3820--9100 \AA.
SEGUE-2, executed during an early stage of the ongoing SDSS-III effort,
observed stars fainter than the SEGUE-1 survey, and added an additional
$\sim$140,000 stars. Stellar spectra obtained as calibration objects or
ancillary projects during regular SDSS-I and SDSS-II operations account
for roughly another 200,000 stars, yielding a total of about 600,000
stars potentially suitable for further analysis. For simplicity, we
refer to all of the SDSS, SEGUE-1, and SEGUE-2 stellar spectra to as
SDSS/SEGUE spectra (stars) throughout this paper.

Since this stellar database includes stars in many evolutionary stages
and spans a wide range of metallicity ($-4.0 <$ [Fe/H] $< +0.5$),
measurement of the carbon-to-iron ratios enables the determination of
the frequency of the CEMP phenomenon for stars of various spectral
types, luminosity classes, and metallicities. This should, in turn,
provide valuable constraints on Galactic chemical evolution models,
numerical simulations of Galaxy formation, the nature of the IMF at
early times, and the various proposed carbon-production mechanisms. 

In this paper, we present new techniques for the measurement of \cfe,
and derive the frequency of CEMP stars from the large sample of
SDSS/SEGUE spectra. Section 2 describes our methodology for estimation
of \cfe. A validation of this method, based on a star-by-star comparison
with high-resolution spectroscopy of SDSS/SEGUE stars in the literature,
is provided in Section 3. The impact of signal-to-noise ratios (S/Ns) on the
measured \cfe\ is also examined in Section 3. Section 4 derives the
frequency of the CEMP stars, as a function of \feh, for the full sample,
as well as for stars in various luminosity classes. A summary of our
results and our conclusions are provided in Section 5.

\section{Measurement of Carbon-to-Iron Ratios ([C/Fe])}

\subsection{Stellar Atmospheric Parameters}

In order to derive the fundamental stellar atmospheric parameters
(\teff, \logg, and \feh) for the SDSS/SEGUE spectra, we employ the most
recent update of the SEGUE Stellar Parameter Pipeline (SSPP; Lee et al.
2008a, 2008b, 2011; Allende Prieto et al. 2008; Smolinski et al. 2011).
The SSPP processes the wavelength- and flux-calibrated SDSS/SEGUE
stellar spectra, and delivers the three fundamental stellar parameters
for most stars with spectral S/N ratio greater than 10 \AA$^{-1}$, in
the temperature range 4000--10,000 K. The SSPP estimates the atmospheric
parameters through a number of approaches, such as a minimum distance
method (Allende Prieto et al. 2006), neural network analysis
(Bailer-Jones 2000; Re Fiorentin et al. 2007), and a variety of
line-index calculations, which were calibrated with respect to known
standard stars (e.g., Beers et al. 1999). This multiple approach permits
the use of as wide a spectral range as possible, in order for the SSPP
to obtain robust estimates of each parameter for stars over a wide range
in \teff, \logg, \feh, and S/N. The SSPP is able to determine \teff,
\logg, and [Fe/H] with typical external errors of 180 K, 0.24 dex, and
0.23 dex, respectively (Smolinski et al. 2011), most reliably for stars
in the temperature range 4500~K $<$ \teff\ $<$ 7500~K. As described by
Lee et al. (2011), the SSPP can also obtain an estimate of
\afe,\footnote[16]{This is generally referred to in the literature as an
average of the abundance ratios [Mg/Fe], [Si/Fe], [Ca/Fe], and
[Ti/Fe].} with a precision of better than 0.1 dex, for SDSS/SEGUE
spectra having S/N $\geq 20$ \AA$^{-1}$.

\subsection{A New Grid of Synthetic Spectra for Determination of \cfe}

Since it is not practical to analyze hundreds of thousands of stellar
spectra one at a time, we have modified the SSPP so that it is capable
of estimating [C/Fe] in a fast, efficient manner. To accomplish this, we
introduce a pre-existing grid of synthetic spectra. This eliminates the
need for generating synthetic spectra on the fly, while simultaneously
attempting to determine the primary atmospheric parameters and/or other
elemental abundances.

As emphasized in Masseron (2006), carbon enhancement affects the
thermodynamical structure of stellar atmospheres. To take into account
this effect, specific models have been tailored with various carbon
abundances using the MARCS code (Gustafsson et al. 2008). From those
models, we have created synthetic spectra using the {\tt Turbospectrum}
synthesis code (Alvarez \& Plez 1998; Plez 2012), which employs the line
broadening treatment described by Barklem \& O'Mara (1998), along with
the solar abundances of Asplund et al. (2005). The sources
of atomic lines used by {\tt Turbospectrum} come from VALD (Kupka et al.
1999), Hill et al. (2002), and Masseron (2006). Line-lists for the
molecular species are provided for CH (T. Masseron et al., in
preparation), and CN and C$_{2}$ (B. Plez, private communication); the
lines of MgH molecules are adopted from the Kurucz
linelists\footnote[17]{http://kurucz.harvard.edu/LINELISTS/LINESMOL/}.

\begin{figure*}
\centering
\plottwo{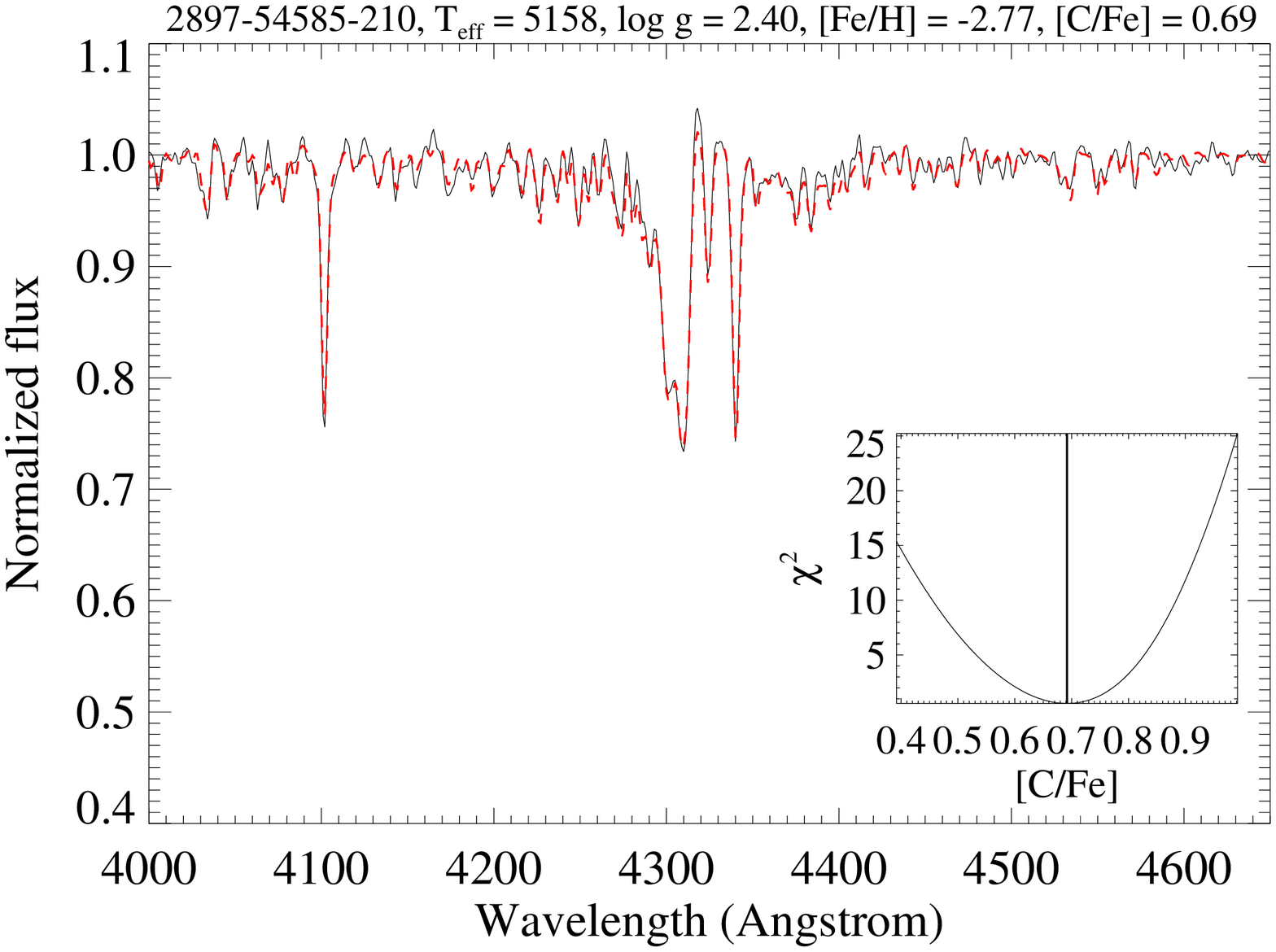}{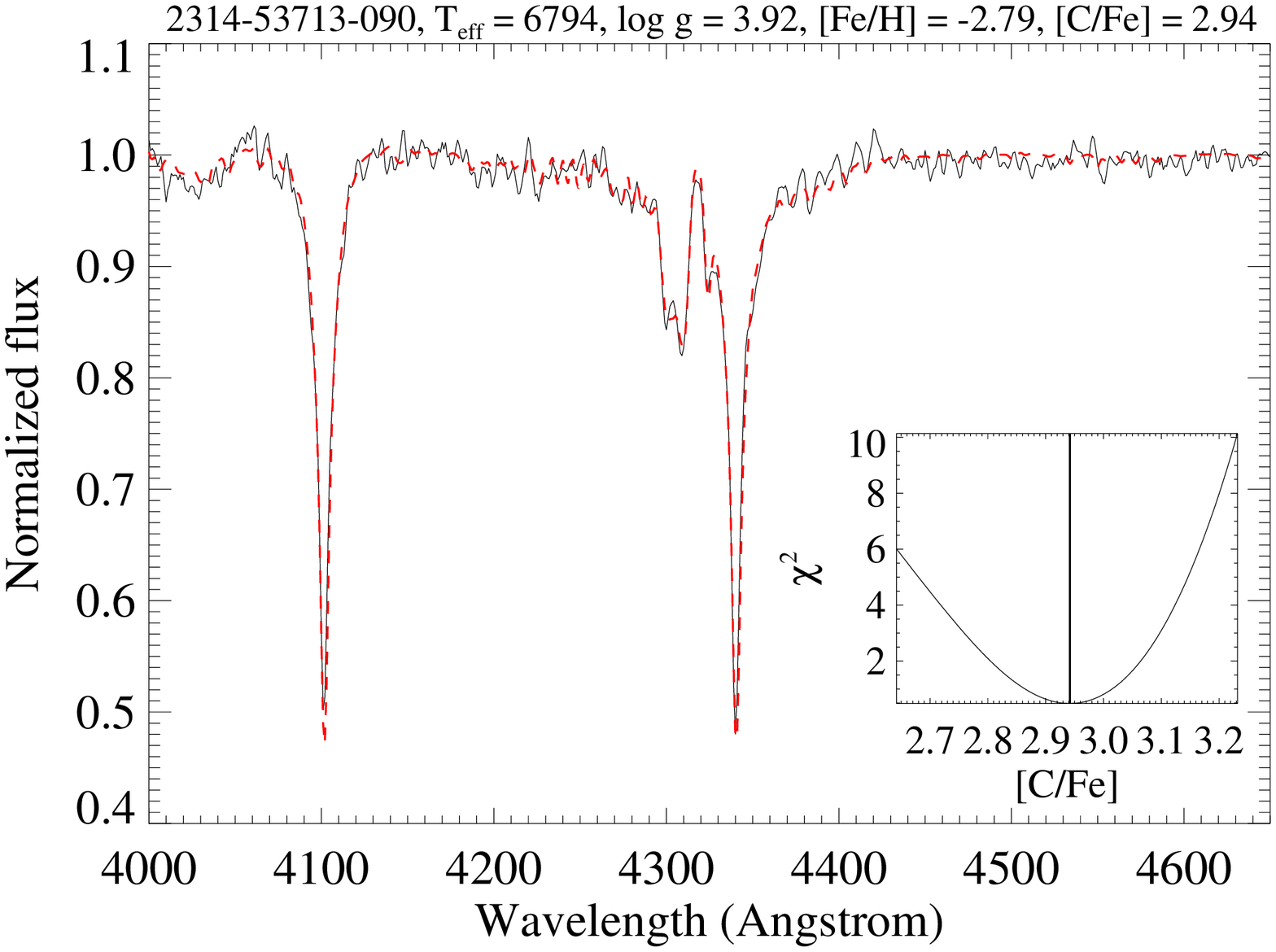}
\caption{Two examples of our spectral matching technique, illustrating
clear detections of the CH $G$-band. The left panel is a cool,
VMP, moderately carbon-enhanced giant ([C/Fe] $= +0.69$), whereas the
right panel exhibits a warm, VMP, highly carbon-enhanced turnoff star
([C/Fe] $= +2.94$). The black line is the observed spectrum, while the
red-dashed line is the best-matching synthetic spectrum generated with
the parameters listed on each plot, as estimated by our approach and
the SSPP. The inset plot in each panel shows the change of the $\chi^{2}$
values over [C/Fe] around the adopted [C/Fe]. The solid vertical
line indicates the measured \cfe\ at the minimum of $\chi^{2}$.}
\label{fig:example}
\end{figure*}

When synthesizing the spectra, we increase (by the same amount) the
abundances for the \alp-elements (O, Mg, Si, Ca, and Ti). We assume that
the $\alpha$-enhancement ratio, relative to Fe, is $+$0.4 for [Fe/H]
$\leq -1.0$, $+$0.2 for [Fe/H] = $-0.5$, and 0.0 for [Fe/H] $\geq 0.0$.
As it is often found that N is enhanced in CEMP stars (Masseron et al.
2010), we assume the same level of N enhancement as for carbon (i.e.,
[N/Fe] = [C/Fe]). We also assumed a carbon isotopic composition of
$^{12}$C/$^{13}$C = 10. We do not consider any neutron-capture element
enhancements (which in any case would have a small effect on spectra at
the resolution of SDSS/SEGUE). In order to assign an appropriate
microturbulence velocity ($\xi_{t}$) for each spectrum, we make use of a
simple polynomial relationship between microturbulence velocity and
surface gravity, $\xi_{t}$ [km s$^{-1}$] = $-0.345\cdot $log $g+2.225$,
derived from the high-resolution spectra of SDSS/SEGUE stars used to
calibrate the SSPP. The synthetic spectra have wavelength steps of 0.01
\AA, covering the wavelength range 4000--5000 \AA, which includes the CH
$G$-band ($\sim$4300 \AA), as well as lines of Sr\,{\sc ii} ($\sim$4077 \AA) and
Ba\,{\sc ii} ($\sim$4554 \AA).

The final grid covers 4000~K $\leq T_{\rm eff} \leq$ 7000~K in steps of
250~K, 1.0 $\leq \log~g \leq $ 5.0 in steps of 0.5 dex, and $-4.0 \leq
\rm [Fe/H] \leq +0.5$ in steps of 0.25 dex. The range of \cfe\ varies
with the metallicity in steps of 0.25 dex as follows:

\begin{itemize}
\item $-0.5 \leq$ [C/Fe] $\leq +3.5$ for [Fe/H] $\leq -1.25$
\item $-0.5 \leq$ [C/Fe] $\leq +1.5$ for $-1.25 <$ [Fe/H] $\leq -0.75$
\item $-0.5 \leq$ [C/Fe] $\leq +1.0$ for [Fe/H] $> -0.75$
\end{itemize}
 
The total number of generated synthetic spectra is 30,069. Once created,
the full set of synthetic spectra are degraded to SDSS resolution
($R=2000$), and re-sampled to 1 \AA~wide linear pixels over the
wavelength 4000--4650 \AA. Each degraded grid is normalized by division
with a pseudo continuum, obtained by the same continuum-fitting routine
used for the observed spectra.

\subsection{Determination of [C/Fe]}

In order to determine \cfe~for the SDSS/SEGUE spectra, we first
transform the vacuum wavelength scale to an air-based scale, and shift
the spectrum to the rest frame using a measured radial velocity. This
wavelength and redshift-corrected spectrum is linearly re-binned to 1
\AA~pixels over the wavelength range 4000--4650 \AA, in which the CH
$G$-band is included. When pre-processing the synthetic spectra, the
spectrum under consideration is then normalized by dividing its reported
flux by its pseudo-continuum. 

The pseudo-continuum over the 4000--4650 \AA~range is obtained by
carrying out an iterative procedure that rejects points lying more than
1$\sigma$ below and 4$\sigma$ above the fitted function, obtained from a
ninth-order polynomial. Although we have a ``perfect'' continuum available
for a given synthetic spectrum, we apply the same continuum routine to
the synthetic and SDSS spectra to match with over the same wavelength
range, and with the same pixel size. Application of the same continuum
routine produces the same level of line-strength suppression in both
spectra.

\begin{figure*}
\centering
\plottwo{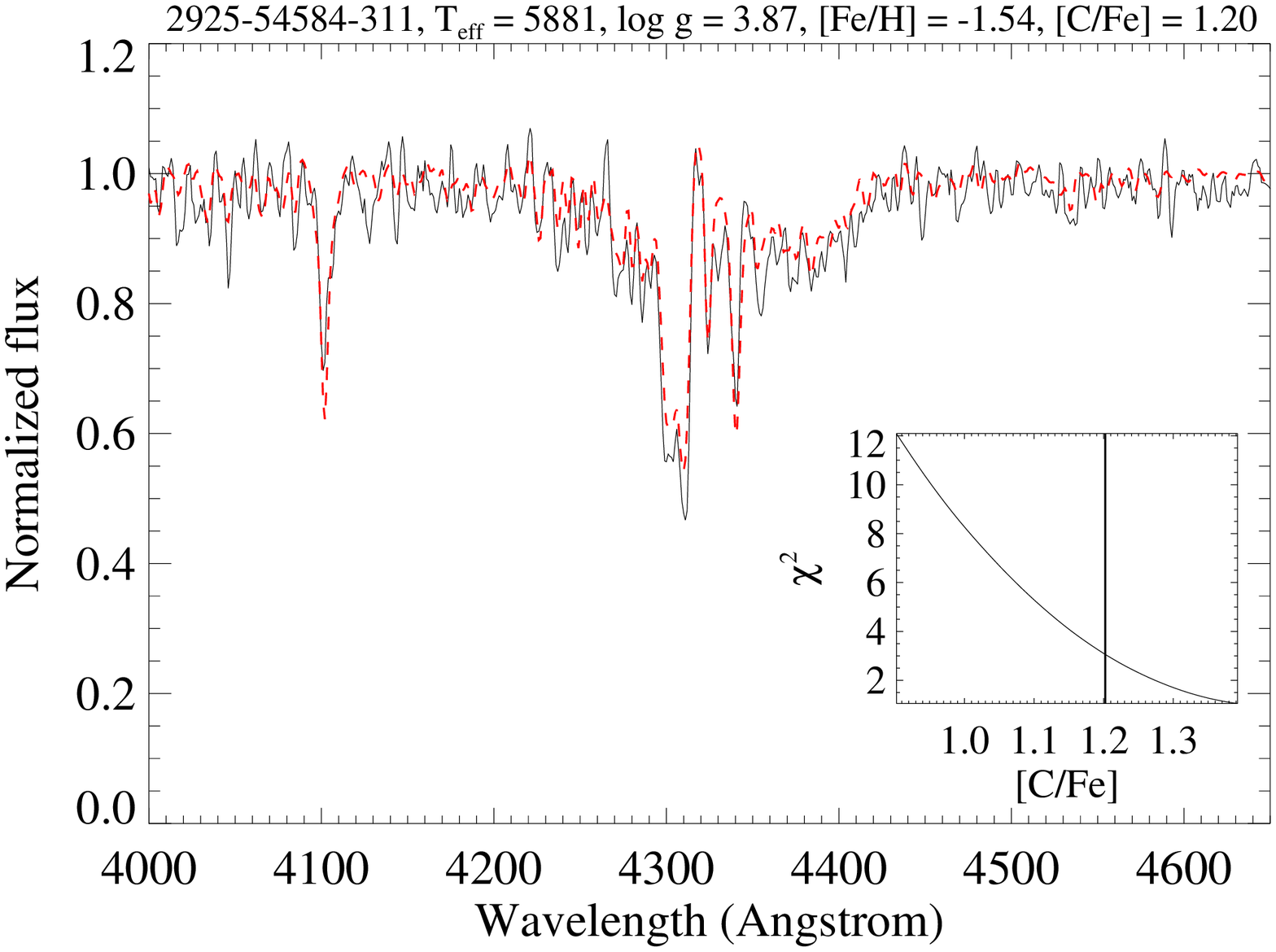}{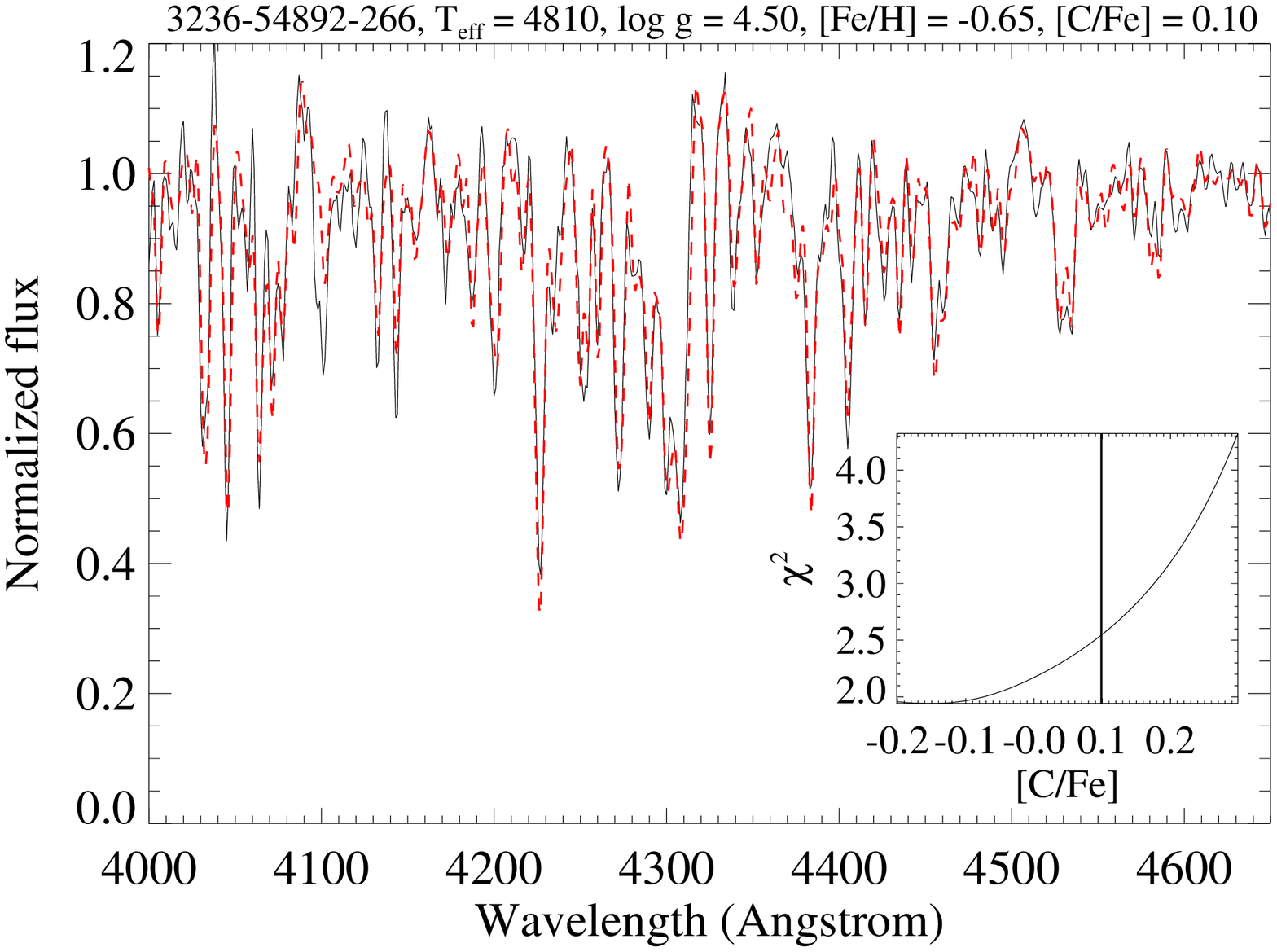}
\caption{Two examples of a lower limit (left panel) and upper limit (right panel) 
on estimates of [C/Fe] from our spectral matching technique. The
spectrum shown in the left panel is a carbon-enhanced ([C/Fe] $= +1.20$),
metal-poor star, while the right panel is a typical, carbon-normal
([C/Fe] = $+0.10$) thick-disk dwarf star. The black line is the observed
spectrum, while the red-dashed line is the best-matching synthetic
spectrum generated with the parameters listed on each plot, as
determined by our methodology and the SSPP. The inset plot in each panel
exhibits the change of $\chi^{2}$ versus [C/Fe]. The solid vertical line
in each panel is where the lower or upper limit is determined (which is
not at the minimum of $\chi^{2}$ values, see text).}
\label{fig:limit}
\end{figure*}

Following the above steps, we then search the grid of synthetic spectra
for the best-fitting model parameters by minimizing the distance between
the normalized target and synthetic flux, using a reduced $\chi^{2}$
statistical criterion. The parameter search over the grid is carried out
by the IDL function minimization routine MPFIT (Markwardt 2009). In this
search, we fix \teff\ and \logg\ to the value determined (previously) by
the SSPP, and change \feh\ and \cfe\ simultaneously to generate a trial
model spectrum by spline interpolation from the existing grid, rather
than vary all four parameters at once. Since the temperature has the
greatest influence on the spectral features over the wavelength range we
consider, holding it constant permits the more subtle variations
associated with the other parameters to be explored. We also find that
the metallicity has a greater impact on measuring [C/Fe] than the
surface gravity. Therefore, we allow [Fe/H] to vary, but not \logg. The
errors in \feh\ and \cfe\ estimated by this approach are determined by
the square root of diagonal elements of the resulting covariance matrix.

Even though the $\chi^{2}$ minimization approach reproduces \cfe\ well,
we seek to improve the accuracy of the carbon measurement so that there
are not spurious detections of carbon-rich stars. Thus, we include an
additional step to check on the estimated [C/Fe]. Briefly, at a given
\teff, \logg, and \feh\ (this metallicity estimate is determined from
the above $\chi^{2}$ minimization), we generate by interpolation a
series of synthetic spectra by varying [C/Fe] by 0.01 dex, from --1.0 to
$+$1.0 dex from the above determined value over the spectral range
4290--4318 \AA\ (in which the prominent CH $G$-band feature exists), and
check how the $\chi^{2}$ values from the differences between the
synthetic spectra and the observed spectrum change. We fit a spline
function to the distribution of the $\chi^{2}$ values against \cfe, in
order to establish the local minimum point where [C/Fe] is determined.
We demand that the extrema of the $\chi^2$ values are at least 10\%
above the local minimum. In this case, we raise a `D' flag, indicative
of a clear detection of CH $G$-band.

In cases where no minimum is found, the $\chi^{2}$ behavior typically
reveals a continuously declining trend of $\chi^{2}$ toward the edge of
the grid. We fit to the portion of the declining function of $\chi^{2}$
with a Gaussian function having a mean value of \cfe\ at the minimum of
declining $\chi^{2}$, in order to estimate the full width half maximum
(FWHM) of this variation. After adding (subtracting) the FWHM to (from)
the \cfe\ value found at the minimum of the declining $\chi^{2}$, we
conservatively define the value as the upper (lower) limit of the
measured \cfe, depending at which edge of the grid the minimum is found.
Accordingly, we raise an `L' flag for a lower limit and a `U' flag for
an upper limit. In this way, we can ensure to have a clear measurement
of the upper or lower limit on [C/Fe], rather than a spurious estimate
of [C/Fe] that can occur in the very low-metallicity (and/or high
\teff) regime, due to the weakness of the CH $G$-band.

It is worth mentioning that the \cfe\ values determined by this approach
agree well with the estimates from the $\chi^{2}$ minimization
technique. We find in most cases small mean offsets ($< 0.05$ dex) and
scatter ($< 0.1$ dex) between the two approaches. 

Figure \ref{fig:example} provides two examples of clear detections of
the CH $G$-band by our spectral fitting method. The left panel is a cool,
VMP, moderately carbon-enhanced giant ([C/Fe] $= +0.69$), while the
right panel shows a warm, VMP, highly carbon-enhanced turnoff star
([C/Fe] $= +2.94$). The black line is the observed spectrum; the
red-dashed line is the best-matching synthetic spectrum generated with
the parameters listed at the top of each plot, as determined by our
methodology and the SSPP. The inset plot in each panel displays how the
$\chi^{2}$ values over 4290--4318 \AA\ change with [C/Fe] around the
adopted [C/Fe], the vertical solid line. From inspection, one can see
that an excellent match between the synthetic and observed spectra is
achieved for these two stars. 

Figure \ref{fig:limit} shows examples of spectra with a lower limit
(left panel) and upper limit (right panel) on \cfe\ obtained following
the technique described above. The spectrum in the left panel is a
carbon-enhanced, metal-poor star, while the right panel is a typical
thick-disk dwarf star. The layout of the figure is the same as for
Figure 1. The inset plots display continuously declining or increasing
values of $\chi^{2}$ over \cfe. The solid vertical line in each panel
indicates where the lower or upper limit is determined from the above
prescription (which is clearly not at the minimum of $\chi^{2}$ values).
As seen in the figure, a lower limit of [C/Fe] $= +1.20$ for the star in
the left panel indicates that it is a highly C-enhanced star, while the
upper limit of \cfe\ $= +0.10$ for the star in the right panel implies
that it is a C-normal star.

\begin{deluxetable*}{cccccrc|ccccccc}[h]
\tablewidth{0in}
\tablecolumns{14}
\setlength{\tabcolsep}{0.001in} 
\tabletypesize{\scriptsize}
\tablecaption{Adopted Atmospheric Parameters and [C/Fe] of the
SSPP and High-resolution Calibration Samples}
\tablehead{\colhead{SDSS Plate--} & \colhead{} &\multicolumn{5}{c}{High Resolution} & \multicolumn{6}{c}{SSPP} & \colhead{} \\
\cline{3-7}  \cline{8-13} 
\colhead{MJD--Fiber} & \colhead{IAU Name} & \colhead{\teff} & \colhead{\logg} & \colhead{\feh} & \colhead{\cfe} & \colhead{$\sigma_{\rm [C/Fe]}$} & 
\colhead{\teff} & \colhead{\logg} & \colhead{\feh} & \colhead{\cfe} & \colhead{$\sigma_{\rm [C/Fe]}$} & \colhead{S/N} & \colhead{Ref.} \\
\colhead{} & \colhead{} & \colhead{(K)} & \colhead{(dex)} & \colhead{(dex)} & \colhead{(dex)} & \colhead{(dex)} & 
\colhead{(K)} & \colhead{(dex)} & \colhead{(dex)} & \colhead{(dex)} & \colhead{(dex)} & \colhead{} & \colhead{}}
\startdata
  0304-51609-528 & SDSS J142237.43$+$003105.2 &   5200 &  2.20 &   $-$3.03 &   $+$1.70 &  0.14 &   5361 &  2.76 &   $-$3.08 &   $+$1.98 &  0.26 & 44 &  A13 \\
  0353-51703-195 & SDSS J170733.93$+$585059.7 &   6700 &  4.20 &   $-$2.52 &   $+$2.10 &  0.32 &   6579 &  3.40 &   $-$2.36 &   $+$2.20 &  0.25 & 62 &  A08 \\
  0471-51924-613 & SDSS J091243.72$+$021623.7 &   6150 &  4.00 &   $-$2.68 &   $+$2.05 &  0.22 &   6211 &  3.38 &   $-$2.64 &   $+$2.44 &  0.25 & 59 &  A13 \\
  0654-52146-011 & SDSS J003602.17$-$104336.3 &   6500 &  4.50 &   $-$2.41 &   $+$2.32 &  0.32 &   6476 &  3.59 &   $-$2.72 &   $+$2.87 &  0.30 & 44 &  A08 \\
  0913-52433-073 & SDSS J134913.54$-$022942.8 &   6200 &  4.00 &   $-$3.24 &   $+$3.01 &  0.22 &   6182 &  3.14 &   $-$3.04 &   $+$3.27 &  0.29 & 40 &  A13 \\
  0938-52708-608 & SDSS J092401.85$+$405928.7 &   6200 &  4.00 &   $-$2.51 &   $+$2.72 &  0.32 &   6239 &  3.31 &   $-$2.25 &   $+$2.65 &  0.28 & 46 &  A08 \\
  0982-52466-480 & SDSS J204728.85$+$001553.8 &   6600 &  4.50 &   $-$2.05 &   $+$2.00 &  0.32 &   6317 &  3.52 &   $-$2.13 &   $+$1.93 &  0.25 & 47 &  A08 \\
  1213-52972-507 & SDSS J091849.91$+$374426.6 &   6463 &  4.34 &   $-$2.98 &   $+$2.82 &\nodata&   6418 &  3.41 &   $-$3.07 &   $+$3.34 &  0.27 & 45 &  Y13 \\
  1266-52709-432 & SDSS J081754.93$+$264103.8 &   6300 &  4.00 &   $-$3.16 &  $<+$2.20 &\nodata&   6111 &  3.28 &   $-$2.85 &   $+$1.31 &  0.26 & 43 &  A08 \\
  1475-52903-110 & SDSS J220924.74$-$002859.8 &   6200 &  4.00 &   $-$3.96 &   $+$2.56 &\nodata&   6539 &  3.55 &   $-$2.87 &   $+$2.44 &  0.28 & 15 &  S13 \\
  1489-52991-251 & SDSS J235718.91$-$005247.8 &   5200 &  4.80 &   $-$3.20 &   $+$0.57 &  0.22 &   5196 &  3.65 &   $-$3.34 &   $+$0.82 &  0.25 & 53 &  A13 \\
  1513-53741-338 & SDSS J025956.45$+$005713.3 &   4550 &  5.00 &   $-$3.31 &   $-$0.02 &  0.22 &   4679 &  4.15 &   $-$3.05 &   $-$0.00 &  0.25 & 43 &  A13 \\
  1600-53090-378 & SDSS J103649.93$+$121219.8 &   5850 &  4.00 &   $-$3.47 &   $+$1.84 &  0.22 &   5949 &  3.19 &   $-$2.82 &   $+$1.55 &  0.28 & 46 &  A13 \\
  1690-53475-323 & SDSS J164610.19$+$282422.2 &   6100 &  4.00 &   $-$3.05 &   $+$2.52 &  0.22 &   6160 &  3.14 &   $-$2.61 &   $+$2.69 &  0.25 & 44 &  A13 \\
  1996-53436-093 & SDSS J112813.57$+$384148.9 &   6449 &  4.38 &   $-$3.53 &  $<+$1.66 &\nodata&   6629 &  3.55 &   $-$2.80 &   $+$0.87 &  0.52 & 45 &  Y13 \\
  2044-53327-515 & SDSS J014036.22$+$234458.1 &   5703 &  3.36 &   $-$4.09 &   $+$1.57 &\nodata&   6155 &  3.51 &   $-$3.64 &   $+$2.03 &  0.26 & 56 &  Y13 \\
  2176-54243-614 & SDSS J161313.53$+$530909.7 &   5350 &  2.10 &   $-$3.33 &   $+$2.09 &  0.14 &   5469 &  2.79 &   $-$2.71 &   $+$1.81 &  0.25 & 51 &  A13 \\
  2178-54629-546 & SDSS J161226.18$+$042146.6 &   5350 &  3.30 &   $-$2.86 &   $+$0.63 &  0.14 &   5418 &  2.58 &   $-$2.60 &   $+$0.39 &  0.25 & 43 &  A13 \\
  2183-53536-175 & SDSS J174624.13$+$245548.8 &   5350 &  2.60 &   $-$3.17 &   $+$1.24 &  0.14 &   5378 &  2.47 &   $-$2.59 &   $+$0.56 &  0.24 & 48 &  A13 \\
  2202-53566-537 & SDSS J162603.61$+$145844.3 &   6400 &  4.00 &   $-$2.99 &   $+$2.86 &  0.22 &   6410 &  3.37 &   $-$2.46 &   $+$2.65 &  0.27 & 41 &  A13 \\
  2309-54441-290 & SDSS J220646.20$-$092545.7 &   5100 &  2.10 &   $-$3.17 &   $+$0.64 &  0.14 &   5167 &  2.12 &   $-$2.94 &   $+$0.50 &  0.24 & 55 &  A13 \\
  2314-53713-090 & SDSS J012617.95$+$060724.8 &   6900 &  4.00 &   $-$3.01 &   $+$3.08 &  0.22 &   6794 &  3.92 &   $-$2.79 &   $+$2.94 &  0.24 & 52 &  A13 \\
  2335-53730-314 & SDSS J030839.27$+$050534.9 &   5950 &  4.00 &   $-$2.19 &   $+$2.36 &  0.22 &   6013 &  3.19 &   $-$2.18 &   $+$2.44 &  0.25 & 41 &  A13 \\
  2337-53740-564 & SDSS J071105.43$+$670228.2 &   5350 &  3.00 &   $-$2.91 &   $+$1.98 &  0.14 &   5421 &  2.32 &   $-$2.67 &   $+$2.22 &  0.25 & 53 &  A13 \\
  2380-53759-094 & SDSS J085833.35$+$354127.3 &   5200 &  2.50 &   $-$2.53 &   $+$0.30 &  0.14 &   5167 &  2.03 &   $-$2.68 &   $+$0.49 &  0.25 & 55 &  A13 \\
  2506-54179-576 & SDSS J114323.43$+$202058.1 &   6240 &  4.00 &   $-$3.15 &   $+$2.75 &\nodata&   6292 &  3.29 &   $-$3.28 &   $+$3.36 &  0.25 & 41 &  S13 \\
  2540-54110-062 & SDSS J062947.45$+$830328.6 &   5550 &  4.00 &   $-$2.82 &   $+$2.09 &  0.22 &   5706 &  3.01 &   $-$2.37 &   $+$2.45 &  0.25 & 46 &  A13 \\
  2552-54632-090 & SDSS J183601.71$+$631727.4 &   5350 &  3.00 &   $-$2.85 &   $+$2.02 &  0.14 &   5361 &  2.40 &   $-$3.09 &   $+$2.76 &  0.27 & 55 &  A13 \\
  2667-54142-094 & SDSS J085136.68$+$101803.2 &   6456 &  3.87 &   $-$2.96 &  $<+$1.39 &\nodata&   6484 &  3.53 &   $-$2.95 &   $+$0.85 &  0.27 & 41 &  Y13 \\
  2679-54368-543 & SDSS J035111.27$+$102643.2 &   5450 &  3.60 &   $-$3.18 &   $+$1.55 &  0.14 &   5631 &  2.93 &   $-$2.77 &   $+$1.81 &  0.24 & 53 &  A13 \\
  2689-54149-292 & SDSS J124123.93$-$083725.5 &   5150 &  2.50 &   $-$2.73 &   $+$0.50 &  0.14 &   5231 &  2.90 &   $-$2.49 &   $+$0.67 &  0.25 & 57 &  A13 \\
  2689-54149-491 & SDSS J124502.68$-$073847.1 &   6100 &  4.00 &   $-$3.17 &   $+$2.53 &  0.22 &   6224 &  3.02 &   $-$2.47 &   $+$2.67 &  0.32 & 41 &  A13 \\
  2799-54368-138 & SDSS J173417.89$+$431606.5 &   5200 &  2.70 &   $-$2.51 &   $+$1.78 &  0.14 &   5421 &  2.12 &   $-$2.75 &   $+$2.55 &  0.29 & 45 &  A13 \\
  2803-54368-459 & SDSS J000219.87$+$292851.8 &   6150 &  4.00 &   $-$3.26 &   $+$2.63 &  0.22 &   6248 &  3.47 &   $-$2.76 &   $+$2.64 &  0.25 & 81 &  A13 \\
  2808-54524-510 & SDSS J170339.60$+$283649.9 &   5100 &  4.80 &   $-$3.21 &   $+$0.28 &  0.22 &   5136 &  3.98 &   $-$3.10 &   $+$0.32 &  0.25 & 82 &  A13 \\
  2857-54453-245 & SDSS J111407.08$+$182831.8 &   6200 &  4.00 &   $-$3.35 &   $+$3.25 &\nodata&   6273 &  3.56 &   $-$3.22 &   $+$3.25 &  0.26 & 42 &  S13 \\
  2897-54585-210 & SDSS J124204.43$-$033618.1 &   5150 &  2.50 &   $-$2.77 &   $+$0.64 &  0.14 &   5158 &  2.40 &   $-$2.77 &   $+$0.69 &  0.24 & 61 &  A13 \\
  2939-54515-414 & SDSS J074104.22$+$670801.7 &   5200 &  2.50 &   $-$2.87 &   $+$0.74 &  0.14 &   5266 &  1.99 &   $-$2.77 &   $+$1.05 &  0.25 & 89 &  A13 \\
  2941-54507-222 & SDSS J072352.21$+$363757.2 &   5150 &  2.20 &   $-$3.32 &   $+$1.79 &  0.14 &   5300 &  2.33 &   $-$3.20 &   $+$2.10 &  0.26 & 59 &  A13
\enddata
\tablecomments{In column labeled Ref. the references are as follows. A08: Aoki et al. (2008); A13: Aoki et al. (2013); Y13: Yong et al. (2013); S13: Spite et al. 2013. Note that the 
stars SDSS J081754.93$+$264103.8, SDSS J112813.57$+$384148.9, and SDSS
J085136.68$+$101803.2 have only reported upper limits on [C/Fe]. S/N is
the average signal-to-noise ratio per Angstrom between 4000 and 8000 \AA\
of the SDSS/SEGUE spectrum. The error estimate of the SSPP \cfe\ follows
from application of Equation1 (\ref{eqn:sys}) and (\ref{eqn:tot}).}
\label{tab:hiressam}
\end{deluxetable*}

\section{Validation of the [C/Fe] Determinations}

\subsection{Star-by-star Comparison with High-resolution Abundance Analysis}

Having developed a new technique for the estimation of [C/Fe], we now
seek to calibrate and validate this method with external measurements.
Although this can be carried out by comparison with the overall behavior
of a sample of stars, it is preferable to compare star-by-star (ideally
against different sources of external measurements), in order to
quantify possible systematic offsets (and optionally remove them), as
well as to determine the likely external errors associated with the
estimate. For these purposes, we employ a set of SDSS/SEGUE stars with
available high-resolution spectroscopy analyzed by Aoki et al. (2008,
2013), Spite et al. (2013), and Yong et al. (2013). Table
\ref{tab:hiressam} lists the stars used to validate our technique, along
with their adopted atmospheric parameters from the high-resolution
spectra, and those used by the SSPP. 

The majority of the stars in Table \ref{tab:hiressam} were analyzed by
Aoki et al. (2008, 2013), who obtained high-dispersion ($R\sim36,000$)
spectra with the Subaru Telescope High Dispersion Spectrograph (Noguchi et al. 2002). 
The four stars listed from Yong et al. (2013) were
observed with Keck/HIRES at $R=48,000$. The spectra for the three stars
from Spite et al. (2013) were collected with VLT/UVES (Dekker et al.
2000) at a resolving power of $R\sim39,000$. Note that, depending on the
adopted model atmospheres, abundance scales, and different effective
temperature scales in the analyses of the high-resolution spectra, there
could be some systematic offsets between these three sets of analyses.
However, we expect any such offsets to be small, and therefore do not
attempt to correct for them here.

\begin{figure*}
\centering
\includegraphics[scale=0.5]{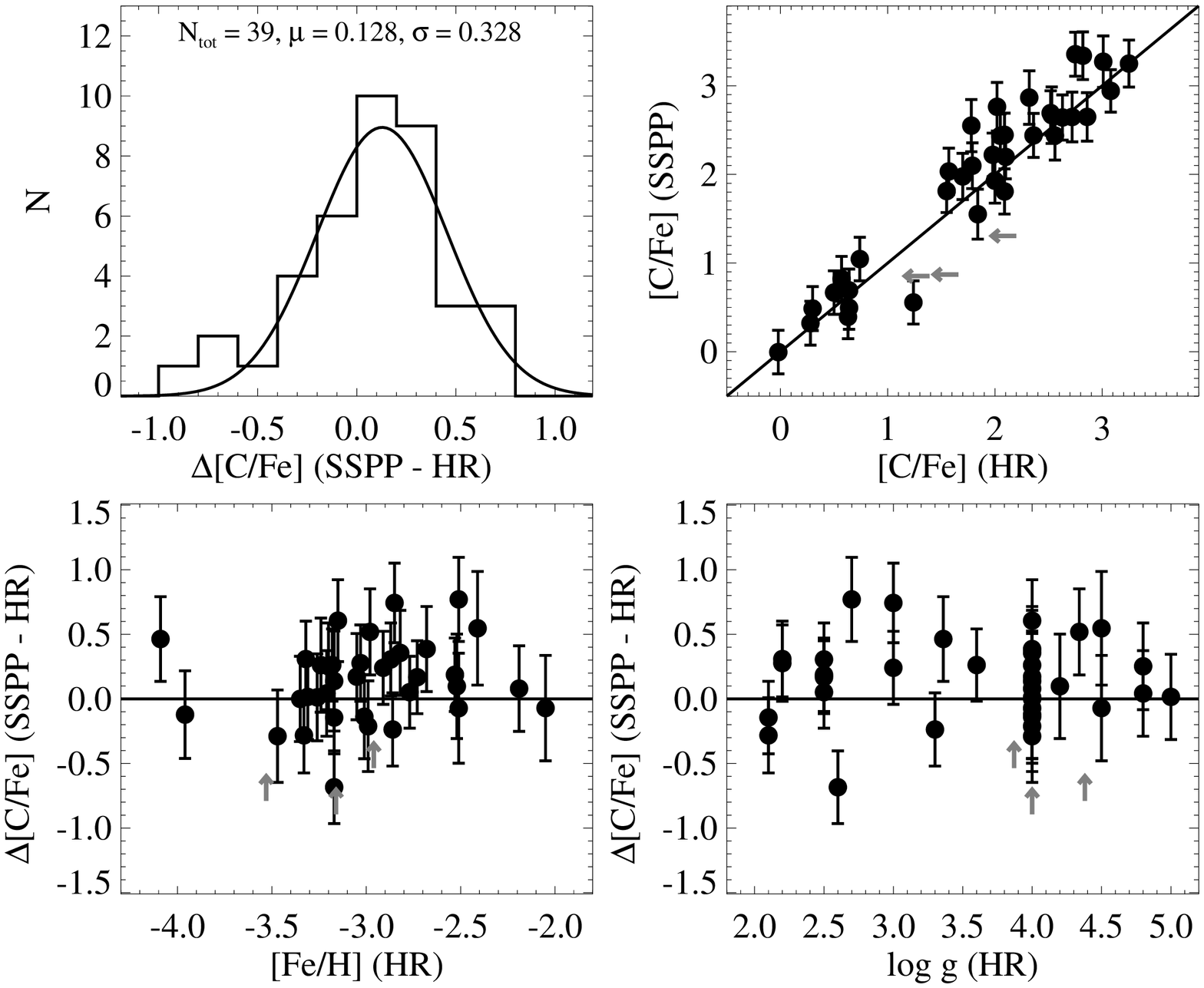}
\caption{Comparisons of our measured carbon abundances (SSPP) with those from the 
literature values based on high-resolution spectroscopy (HR), as listed in 
Table \ref{tab:hiressam}. A systematic offset of 0.128 dex, with a
standard deviation of 0.328 dex, is noted from the Gaussian fit to the
differences between the two estimates shown in the upper-left panel. The
left-pointing arrows in the upper-right panel, and the upward pointing
arrows in the lower two panels, indicate upper limit estimates 
on \cfe\ delivered from the literature (they are reported as
detections with lower values from the SSPP). The error bars in the two
lower panels are obtained from the quadratic addition of our measured
total error to the literature error, while the error bars in the
upper-right panel are the total errors computed from application of
Equation (\ref{eqn:tot}).}
\label{fig:hicfe}
\end{figure*}

One limitation of this comparison sample is that, as noted from
inspection of the table, the sample consists of mainly main-sequence
turnoff stars with \feh\ $< -2.0$; hence it does not cover a wide range
of \teff\ and \feh. Nevertheless, since it is more important to obtain
accurate \cfe\ estimates for metal-poor stars than metal-rich stars, it
should still serve as an excellent sample to test how well our method
performs in the low-metallicity regime.

Figure \ref{fig:hicfe} illustrates a comparison of our measured
carbon-abundance ratios with the adopted literature values based on
high-resolution abundance analyses. The notation ``SSPP'' refers to the
analysis of the low-resolution SDSS/SEGUE spectra, while ``HR'' denotes
the high-resolution determination. An average systematic offset of 0.128
dex, with a standard deviation of 0.328 dex, is noticed from the
Gaussian fit to the residuals (SSPP$-$HR) between the two estimates,
shown in the upper-left panel. Note that this offset is not unexpected,
given the use of our improved carbon-enhanced model atmospheres
(carbon-enhanced models are not used in most high-resolution abundance
analyses due to the difficulty of generating such models), and the
different molecular linelists employed by the high-resolution analyses.
As the offset is small, and in any case much lower than the rms scatter,
we do not attempt to adjust our measurements. There are also no strong
trends with either [Fe/H] or \logg, as shown in the lower panels. The
left-pointing arrows in the upper-right panel, and the upward pointing
arrows in the lower two panels, indicate upper limit estimates on \cfe\
drawn from the literature (they are reported as detections with lower
values from the SSPP).

As the uncertainty in our measured \cfe~includes both external and
internal random errors, and the abundances that we employ from the literature
(based on the high-resolution analyses) carries its own errors as well,
we quantify the total error in our measurement of \cfe~by the following
procedure. Let $\sigma_{\rm g}$ be the rms scatter (e.g., 0.328 dex in
Figure \ref{fig:hicfe}) from a Gaussian fit to the differences between
our measurements and the literature values of \cfe, and $\sigma_{i,\rm
HR}$ be the reported error from the $i$th star in the literature sample (LS).
Then, the external error ($\sigma_{i,\rm ext}$) in our measured \cfe\
of the $i$th object is derived by: 

\begin{equation}
\sigma_{i,\rm ext} = \sqrt{\sigma_{\rm g}^{2} - \sigma_{i,\rm HR}^{2} - \sigma_{i,\rm SSPP}^{2}},
\label{eqn:sys}
\end{equation}

\noindent where $\sigma_{i,\rm SSPP}$ is the random error of the SSPP 
determination for the $i$th entry, simply taken to be the internal
uncertainty of our technique, estimated by following the procedure
described in Section 2.3 above. In this equation, as the subscript
indicates, $\sigma_{i,\rm HR}$ and $\sigma_{i,\rm SSPP}$ are based on
the individual values for each target, whereas a value of $\sigma_{\rm
g}$ from the full sample is used. In other words, when calculating
$\sigma_{i,\rm ext}$, $\sigma_{\rm g}$ is fixed for all objects, while
$\sigma_{i,\rm HR}$ and $\sigma_{i,\rm SSPP}$ change for each star.
Table \ref{tab:hiressam} indicates there are eight stars without
reported error estimates. For these stars, we adopt the average of the
errors from other stars as their associated error. We take an average of
the errors in the literature as well as in the SSPP to derive the
overall average external error, defined as $\langle\sigma_{\rm ext}\rangle$. We
obtain $\langle\sigma_{\rm ext}\rangle$ = 0.221 dex from Equation (\ref{eqn:sys}).
This $\langle\sigma_{\rm ext}\rangle$ is applied to individual estimates of the
SSPP-derived \cfe\ to yield the total error in our measurement
of \cfe~for each object by the following equation:

\begin{equation}
 \sigma_{i,\rm tot} = \sqrt{\langle\sigma_{\rm ext}\rangle^{2} +~\sigma_{i,\rm SSPP}^{2}}.
\label{eqn:tot}
\end{equation}

In the equation above, the largest contribution to the total error
($\sigma_{i,\rm tot}$) comes from the scatter ($\sigma_{\rm g}$) between
our measured values and the literature values. However, if the noise
increases in a given spectrum (for instance, from a low S/N spectrum),
both the external and random errors contribute more to the total
error. 

\begin{figure}
\centering
\plotone{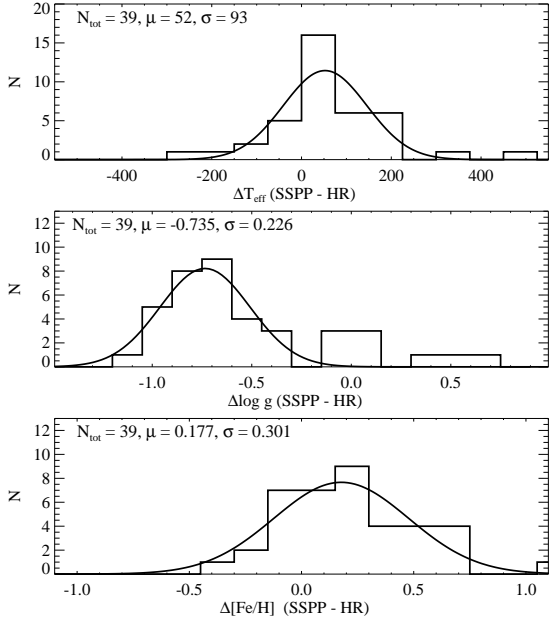}
\caption{Comparisons of \teff, \logg, and \feh\ from the SSPP with those from 
the literature values based on high-resolution spectroscopy (HR), as
listed in Table \ref{tab:hiressam}. There is a small offset with a
small scatter for the effective temperature, as seen in the top panel.
However, there appears to exist a large offset in \logg, evident in the middle
panel---see text for a discussion of the cause of this apparent offset, and 
reasons for believing that it is in actuality much smaller. The 
routine for estimating \cfe\ tends to overestimate the metallicity by 
about 0.18 dex for this sample of metal-poor stars, as seen in the bottom panel.}
\label{fig:hipar}
\end{figure}

The error bars in the lower panels of Figure \ref{fig:hicfe} are
obtained from the quadratic addition of our measured total error and the
literature error, whereas the error bars in the upper-right panel are the
total errors computed by Equation (\ref{eqn:tot}). Note that, as mentioned
earlier, there are some stars without properly measured errors in the
literature values. For those stars, we adopt an average \cfe~uncertainty
based on all stars with available error estimates. We do not take into
account the small mean offset (0.128 dex) in our total error
calculations. The quoted total errors of the SSPP \cfe\ in Table
\ref{tab:hiressam} are calculated from Equations (\ref{eqn:sys}) and
(\ref{eqn:tot}) above, and are mostly less than 0.3 dex.

We also compare the atmospheric parameters (\teff, \logg, and \feh) from
the high-resolution analyses with those obtained from the SSPP; Figure
\ref{fig:hipar} exhibits the results of the comparisons. Note that the
temperature and gravity come directly from the SSPP, while the
metallicity is estimated during the $\chi^{2}$ minimization. We notice 
a small offset in \teff, of 52~K, with $\sigma = 93$~K,

There appears to be a large systematic offset, of about 0.7
dex, in the gravity estimate. There could be several reasons for this. 
One is that the gravity estimates by Aoki et al. (2008, 2013) 
in Table \ref{tab:hiressam} are based on snapshot high-resolution 
spectra, which have generally S/N less than 50 per resolution element. 
Some of these stars, those warmer than 5500 K, were assumed by Aoki et al. 
to have \logg\ = 4.0 (as they are turnoff stars) because of their very 
weak Fe\,{\sc ii} features precluded using them to estimate gravity 
by the usual procedure of forcing the iron abundance from the Fe\,{\sc i} 
lines to match with that derived from the Fe\,{\sc ii} lines. If an accurate estimate 
of surface gravity were possible to obtain from these data, it would be expected to 
be between 3.5 and 4.5. In addition, three stars cooler than 5200 K 
from Aoki et al. (2013) were classified as main-sequence stars 
on the basis of strong strength of their Mg\, {\sc i} b lines and weak features of 
ionized atoms. The gravity for these stars was estimated from 
matching theoretical isochrones for old, metal-poor dwarf stars; two 
stars with \logg\ = 4.8 and one with \logg\ = 5.0 were claimed.

If we set aside the above stars (as well as three stars from Spite et al. 2013 
that were also assigned gravities of \logg\ = 4.0, rather than having 
their surface gravity derived), we are left with 21 stars for which gravity 
estimates were obtained based on the analysis of Fe\,{\sc i} and Fe\,{\sc ii} 
lines. The distribution in the residuals in the gravity between the SSPP and the 
high-resolution literature values for these objects is then too broad 
to derive a Gaussian mean and scatter (and the sample size is not 
sufficiently large to derive meaningful Gaussian statistics). Instead, taking 
the simply arithmetic mean, the derived zero-point offset for this sample is 
$<\Delta$(\logg)$>$ = --0.3 dex, with a standard deviation of 0.5 dex, a much 
smaller offset with a slightly larger scatter than derived from the full sample.

Another likely contribution to the apparently large offset in surface 
gravity is the fact that the calibration sample primarily comprises VMP ([Fe/H] $< -2.5$) 
and relatively warm stars, which exhibit very 
weak gravity-sensitive features at the low resolution of the SDSS/SEGUE 
spectra. A sample of 126 SDSS/SEGUE stars with parameters 
based on high-resolution analyses has recently been used to re-calibrate 
the SSPP (C. Rockosi et al., in preparation). This analysis justifies the above
claim, as Rockosi et al. report obtaining a mean zero-point offset of 
$\langle\Delta$(\logg)$\rangle$ = 0.0 dex, with an rms scatter of 0.4 dex, 
from 67 stars with [Fe/H] $> -2.5$ and \teff\ $> 5100$ K (all of the stars 
in Table \ref{tab:hiressam} are hotter than 5100 K), whereas they 
obtain a mean zero-point offset of $\langle\Delta$(\logg)$\rangle$ = --0.3 dex, with a 
scatter of 0.5 dex from 26 stars with [Fe/H] $< -2.5$ and \teff\ $> 5100$ K, 
the same result found from the trimmed sample of 21 stars discussed above. 
Based on these results, we conclude that a more realistic estimate of 
the systematic offset in the SSPP gravity estimate (for VMP, 
warm turnoff stars) is on the order of about 0.3 dex rather than 
about 0.7 dex. This offset is smaller than the rms scatter, and essentially 
corresponds to the uncertainty in this parameter that one can derive 
from high-resolution spectroscopy. The above results also suggest that the SSPP 
gravity estimates for relatively more metal-rich stars should 
be more accurate, since the features of the gravity-sensitive lines are 
much stronger in such stars.

As seen in Figure \ref{fig:hicfe}, we have obtained a small offset
(about 0.1 dex) in \cfe\ between the SSPP and the high-resolution
analyses, implying a good agreement between these estimates.
Furthermore, our error analysis in Table \ref{tab:hiressam} indicates a
typical error of 0.3 dex in our measured \cfe, which is relatively small
given that the high-resolution analysis can also produce the error on
the order of 0.2 dex. This indicates that the small systematic deviation 
in the estimate of \logg\ from the SSPP in this EMP regime does not 
strongly influence our determination of \cfe. In the 
following subsection, additional tests of the effects of the parameter 
errors from the SSPP on the measured \cfe\ are discussed. 

Figure \ref{fig:hipar} also indicates that our estimate of \feh\ is
slightly higher, by 0.18 dex, in this VMP regime (\feh\ $\leq
-2.5$), with a scatter of about 0.30 dex. Even if the comparison results
suggest some systematic offsets in all three parameters, we do not
attempt to correct the offsets when interpreting the C-rich stars, as
they are relatively small and do not greatly influence 
(in the case of \logg) our measurement of \cfe.

We have decided to use the adopted \feh\ from the SSPP, not the
one from the $\chi^{2}$ minimization obtained while estimating \cfe, as
it exhibits a smaller offset (0.08 dex) and scatter (0.16 dex) when
compared to the high-resolution results. In order to account for the
subtle change of \cfe\ owing to the use of the adopted \feh, we
recalculate [C/Fe]$_{\rm adjusted}$ by [C/H] -- [Fe/H]$_{\rm adopted}$,
where [C/H] comes from [C/Fe] $+$ [Fe/H] from the carbon-determination
routine. Hence, in our analysis of C-enhanced stars, we make use of
[C/Fe]$_{\rm adjusted}$ and [Fe/H]$_{\rm adopted}$, and simply report
these values as \cfe\ and \feh.

\subsection{Effects of Errors in \teff\ and \logg\ on Determination of \cfe}

During the process of carrying out the minimum $\chi^{2}$ search, we
have fixed \teff\ and \logg\ at the values determined by the SSPP, and
only allow the other two parameters, \feh\ and \cfe, to be solved for
simultaneously. However, since the effective temperature and surface
gravity estimates delivered by the SSPP themselves carry uncertainties,
we need to examine how the errors in \teff\ and \logg\ propagate into
uncertainties in the determination of \cfe\ and \feh. We perform this
test (using the high-resolution validation sample) by varying the
adopted \teff~by $-$300, $-$200, $-$100, $+$100, $+$200, and $+$300~K,
and $-$0.7, $-$0.5, $-$0.3, $+$0.3, $+$0.5, and $+$0.7~dex for the 
adopted gravities suggested by the SSPP. The deviation of 0.7 dex is 
the amount found from the comparison with the high-resolution LS 
in Figure \ref{fig:hipar}. 

\begin{deluxetable*}{cccccccccc}
\centering
\tablewidth{0pc}
\setlength{\tabcolsep}{0.001in} 
\tabletypesize{\scriptsize}
\tablecaption{Effects of Errors in \teff\ and \logg\ on Determination of \feh\ and \cfe}
\tablehead{\colhead{\teff} & \multicolumn{2}{c}{\feh} & \multicolumn{2}{c}{\cfe} & \colhead{\logg} & \multicolumn{2}{c}{\feh} & \multicolumn{2}{c}{\cfe} \\
\cline{2-3}  \cline{4-5}  \cline{7-8}  \cline{9-10}
\colhead{Error} & \colhead{$\mu$} & \colhead{$\sigma$} & \colhead{$\mu$} & \colhead{$\sigma$} & \colhead{Error} & \colhead{$\mu$} & \colhead{$\sigma$} & \colhead{$\mu$} & \colhead{$\sigma$}\\
\colhead{(K)} & \colhead{(dex)} & \colhead{(dex)} & \colhead{(dex)} & \colhead{(dex)} & \colhead{(dex)} & \colhead{(dex)} & \colhead{(dex)} & \colhead{(dex)} & \colhead{(dex)}}
\startdata
$-$300  &  $-$0.201  &  0.352  &  $-$0.391  &  0.277  &  $-$0.7  &  $-$0.063  &  0.304  &  $+$0.266  &  0.349   \\
$-$200  &  $-$0.173  &  0.317  &  $-$0.234  &  0.254  &  $-$0.5  &  $-$0.069  &  0.274  &  $+$0.185  &  0.332   \\
$-$100  &  $-$0.123  &  0.277  &  $-$0.110  &  0.232  &  $-$0.3  &  $-$0.066  &  0.253  &  $+$0.153  &  0.280   \\
  $+$0  &  $+$0.000  &  0.283  &  $-$0.005  &  0.206  &  $+$0.0  &  $+$0.000  &  0.283  &  $-$0.005  &  0.206   \\
$+$100  &  $+$0.065  &  0.294  &  $+$0.167  &  0.259  &  $+$0.3  &  $+$0.024  &  0.339  &  $-$0.104  &  0.332   \\
$+$200  &  $+$0.086  &  0.346  &  $+$0.256  &  0.291  &  $+$0.5  &  $+$0.020  &  0.335  &  $-$0.176  &  0.359   \\
$+$300  &  $+$0.132  &  0.308  &  $+$0.404  &  0.317  &  $+$0.7  &  $+$0.030  &  0.397  &  $-$0.234  &  0.378   
\enddata
\tablecomments{The symbol $\mu$ is the Gaussian mean in the residuals between the SSPP and the high-resolution values, while $\sigma$ 
is calculated following Equations (\ref{eqn:sys}) and (\ref{eqn:tot}). These are derived after adjusting 
for offsets of 0.177 dex for \feh, and 0.128 dex for \cfe, found in Figures \ref{fig:hipar} and \ref{fig:hicfe}, respectively.}
\label{tab:error}
\end{deluxetable*}

Table \ref{tab:error} summarizes the results of this experiment, and
lists the derived variations in the estimated \feh\ and \cfe. As we are
primarily interested in how the shifts in the temperature and gravity
affect the \feh\ and \cfe\ estimates, we first remove the offsets of
0.177 dex and 0.128 dex for \feh\ and \cfe, respectively, found in
Figures \ref{fig:hipar} and \ref{fig:hicfe}. Then, we derive the
mean offsets and standard deviations from Gaussian fits to the residual
distributions. The middle row of the table lists the mean and scatter
after removing the offsets (without applying the \teff\ and 
\logg\ shifts). Note that the scatter ($\sigma$) in the table is similarly 
calculated by following Equations (\ref{eqn:sys}) and (\ref{eqn:tot}). To calculate the
external error for \feh, we assume the error in the literature values to
be 0.15 dex in \feh, as the typical uncertainty of the high-resolution
analysis is 0.1--0.2 dex. 

Starting with the temperature shifts, even when the temperature is
systematically deviated by 100 K, 200 K, or 300 K, we do not notice much
change (less than 0.11~dex at most) in the rms scatter for both \feh\
and \cfe. We do see, however, variations in the offsets up to $\sim$
0.2 dex in \feh, and about 0.4 dex in \cfe, at the most extreme shift of
\teff. For a shift of 200 K, which is the same order of magnitude of the
error in \teff\ from the SSPP, the deviation is less than 0.25 dex for
both \feh\ and \cfe, smaller than the rms scatter. Therefore, unless the
SSPP \teff\ is grossly incorrect, the \feh\ and \cfe\ estimates may not
systematically change by more than 0.25 dex.

Table \ref{tab:error} also suggests that the Gaussian scatter of \feh\
and \cfe\ does not vary much from shifting \logg, with much lower offsets
in \feh\ and \cfe\ than for the \teff\ shifts. Once again, this 
confirms the above claim that the gravity error in the SSPP has only a 
very minor impact on our measured parameters. This is a very encouraging 
result, in that the systematic error in the gravity estimate from 
the SSPP can be not only as large as 0.5 dex at the base of the red 
giant branch, and as high as about 0.3 dex for warm, EMP stars as 
found in the previous section.

Summarizing, for the two derived parameters (\feh\ and \cfe) the mean
offsets associated with different input offsets in \teff\ and \logg\ are
mostly smaller than the derived rms scatter in the determinations of
these parameters. Accordingly, it appears that, within $\pm$200~K, which
is equivalent to the typical error of the SSPP-determined \teff\ 
(Smolinski et al. 2011), the \cfe~estimate is perturbed by less than
$\pm$0.25 dex, which is smaller than the rms scatter of 0.30~dex, the
typical total error of our measured \cfe\ from Table \ref{tab:hiressam}.
This implies that our approach to deriving \cfe~is robust against 
small (systematic) deviations of the SSPP-derived temperature and
surface gravity.

\subsection{The Impact of Signal-to-noise Ratios on Parameter Estimates}

\subsubsection{Noise-added Synthetic Spectra}

Because they are relatively bright, the stars with
high-resolution estimates of [C/Fe] used to validate our technique
typically have high S/N ($>$40) SDSS/SEGUE spectra. However, since the
full sample of SDSS/SEGUE spectra covers a wide range of S/N, it is
desirable to check how declining S/N impacts our estimation of \cfe. 

To test this, we first inject noise (which mimics the uncertainty in the
flux of an SDSS/SEGUE spectrum) into the grid of synthetic spectra used
to estimate the  carbon abundance ratio. We select a list of the SDSS/SEGUE spectra
that have similar \teff\ and \feh\ to the synthetic spectrum to which we
wish to add noise. From the selected SDSS/SEGUE spectra, we choose a
spectrum that has an average S/N value to target for the noise-injected
synthetic spectrum. Using the observed S/N values as a function of
wavelength, we generate a noise array by dividing the synthetic flux by
the S/N values of the selected SDSS/SEGUE spectrum. We then add this
noise to the synthetic spectrum by assuming that the noise is a
1$\sigma$ error of a Gaussian distribution. Through the same process, we
introduce noise-added synthetic spectra having S/N = 7.5, 10.0, 12.5,
15.0, 20.0, 25.0, 30.0, 35.0, 40.0, 45.0, and 50.0 \AA$^{-1}$. We
construct 25 different noise-added synthetic spectra at each S/N value.
These noise-added spectra are processed following the same procedures to
determine [C/Fe]. In that process, we hold \teff\ and \logg\ values
associated with the synthetic spectra constant, and change \feh\ and
\cfe\ to minimize the $\chi^{2}$ values.

\begin{figure*}
\centering
\includegraphics[scale=0.8]{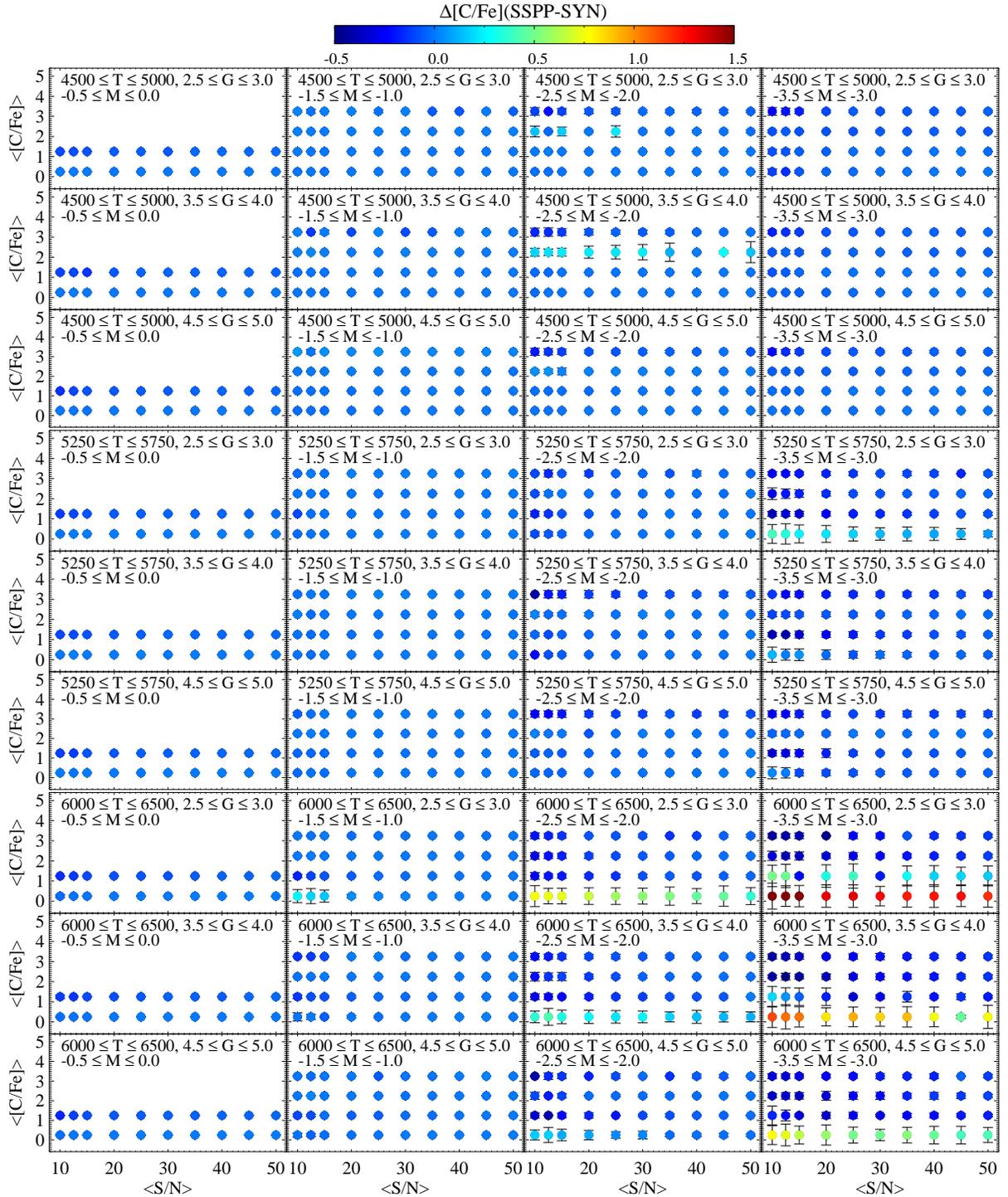}
\caption{Distributions of residuals in [C/Fe] between the SSPP and
synthetic model values. $\langle$S/N$\rangle$ is an average signal-to-noise ratio 
per Angstrom, whereas $\langle$[C/Fe]$\rangle$ is an average value of the models with \cfe\ = 0.0
to $+$0.5, $+$1.0 to $+$1.5, $+$2.0 to $+$2.5, and $+$3.0 to $+$3.5,
that is, $\langle$[C/Fe]$\rangle$ = $+$0.25, $+$1.25, $+$2.25, and $+$3.25,
respectively. The color-coded circles denote the difference in \cfe; the
scale of the differences is displayed as a color bar at the top. `SSPP'
denotes the SSPP values, while `SYN' indicates the model values. The
temperature (T), gravity (G), and metallicity (M) ranges are listed 
at the top of each panel.}
\label{fig:syntest}
\end{figure*}

To examine how the S/N affects the estimation of [C/Fe], we group the
noise-added synthetic spectra into three ranges of \teff\ (4500~K to
5000~K, 5250~K to 5750~K, and 6000~K to 6500~K), three ranges of \logg\
(2.5 to 3.0, 3.5 to 4.0, and 4.5 to 5.0), four regions of \feh\ (0.0 to
--0.5, --1.0 to --1.5, --2.0 to --2.5, and --3.0 to --3.5), and four
regions of \cfe\ (0.0 to $+$0.5, $+$1.0 to $+$1.5, $+$2.0 to $+$2.5, and
$+$3.0 to $+$3.5). Then, we derive the Gaussian mean and sigma of the
differences in \cfe\ between the SSPP-estimated values and the model
values for a group of spectra that fall within the combination of the
parameter ranges at each S/N value. Figure \ref{fig:syntest} shows how
the mean and scatter change with the average S/N at different levels of
carbon enhancements ($\langle$[C/Fe]$\rangle$). In the figure, $\langle$[C/Fe]$\rangle$ 
is an average value of the models with [C/Fe] = 0.0 to $+$0.5, $+$1.0 to
$+$1.5, $+$2.0 to $+$2.5, and $+$3.0 to $+$3.5, that is, $\langle$[C/Fe]$\rangle$ =
$+$0.25, $+$1.25, $+$2.25, and $+$3.25, respectively. The error bars
indicate the Gaussian scatter, and are mostly smaller than the symbol
size. The color-coded circles denote the difference in [C/Fe], and the
scale of the differences is displayed as a color bar at the top of the
plot. The label `SSPP' denotes the SSPP values, while `SYN' indicates
the model values. The temperature (T), gravity (G), and metallicity (M)
ranges are indicated at the top of each panel.

From inspection of Figure \ref{fig:syntest} it is clear that, over most
of the parameter regions, there are no large deviations in the zero
point or scatter around the mean as a function of S/N for different 
carbon enhancements. However, as the metallicity decreases and temperature 
increases, the circle symbols turn from blue (indicating a small
systematic offset in \cfe) to yellow and red (indicative of a larger
systematic offset in \cfe), becoming quite severe for the lowest gravity
and lowest [C/Fe] model spectra. The primary reason for this 
is that the CH $G$-band strength does not change much with varying carbon
abundance in these low-metallicity, high-temperature, and especially
C-normal (\cfe\ $< +1.0$) regimes. This effect becomes worse for the
low-gravity ranges (\logg\ $ < 3.0$); the strength of the CH $G$-band is
almost indistinguishable among the models with different carbon
abundances in the high-temperature, low-gravity, low-metallicity, and
low-carbon abundance ranges. This causes a very broad and poorly
constrained distribution of $\chi^{2}$ values as a function of \cfe,
which results in unreliable estimate of \cfe\ with much larger error ($>
0.5$ dex). As a result, the low carbon-abundance models significantly deviate
from the zero point in this parameter range, as seen in Figure
\ref{fig:syntest}. However, we do not expect to see many stars in the
high-temperature (\teff\ $> 6000$ K), low-gravity (\logg\ $< 3.0$)
ranges in our SDSS/SEGUE sample (Presumably they would be red
horizontal-branch stars, not main-sequence turnoff stars. See Section
4.1).

Except for this concern, it is clear that our technique reproduces \cfe\
reasonably well, as the rms scatter between the SSPP analysis and the
models are mostly less than 0.3 dex, without significant deviation in
the zero points, over most of the parameter space down to S/N = 15
\AA$^{-1}$.

\subsubsection{Noise-added SDSS/SEGUE Spectra with High-resolution Parameters}

We perform another noise-injection experiment for the spectra of stars
listed in Table \ref{tab:hiressam}, following the same prescription
described above. As we are not able to boost the S/N to values larger
than the original S/N of a spectrum, we only generate the noise-added
spectra up to a level below the original S/N of the SDSS/SEGUE spectrum.
For example, the first entry (SDSS J142237.43+003105.2) in Table
\ref{tab:hiressam} has S/N = 44 \AA$^{-1}$, and we create
noise-injected spectra up to S/N = 40 \AA$^{-1}$. Thus, for this
spectrum, we generate spectra with S/N = 7.5, 10, 12.5, 15, 20, 25, 30,
35, and 40 \AA$^{-1}$. As before, we produce 25 different realizations
at each S/N value. The same procedure is applied to other stars. The
star SDSS J220924.74$-$002859.8, with S/N = 15 \AA$^{-1}$, only has
noise-injected spectra up to S/N = 15 \AA$^{-1}$.

Once generated, the noise-injected spectra are processed through the
SSPP to obtain estimates of \teff, \logg, \feh, and \cfe, and we examine
how each parameter changes with S/N. Table \ref{tab:noise} summarizes
the results of the experiment. The Gaussian mean ($\mu$) and sigma
($\sigma$) are calculated from considering all spectra that fall in each
S/N bin (e.g., 39$\times$25 = 975 spectra for S/N = 10 \AA$^{-1}$),
after adjusting for the offsets of 52~K for \teff, --0.735 dex for \logg,
0.177 dex for \feh, and 0.128 dex for \cfe\ (which are found in Figures
\ref{fig:hipar} and \ref{fig:hicfe}, respectively). 

Inspection of Table \ref{tab:noise} indicates that the magnitude of the
mean offsets and rms scatters generally increase with declining S/N for
\teff, \logg, \feh, and \cfe\, as expected. Note that the rms scatters are calculated
by following Equations (\ref{eqn:sys}) and (\ref{eqn:tot}). We assume errors
of 0.3 dex in \logg\ and 0.15 dex in \feh\ for the high-resolution
estimates, whereas we do not take into account the error in \teff\ from
the high-resolution results. One can notice from the table a very small
offset in \teff, with scatter up to 170 K, relative to the
high-resolution values, as the S/N decreases. 

\begin{deluxetable}{crrcccccc}
\centering
\tablewidth{0pc}
\setlength{\tabcolsep}{0.001in}
\tabletypesize{\scriptsize}
\tablecaption{Impact of Signal-to-noise Ratios on Determination of \teff, \logg, \feh, and \cfe}
\tablehead{\colhead{} & \multicolumn{2}{c}{\teff} & \multicolumn{2}{c}{\logg} & \multicolumn{2}{c}{\feh} & \multicolumn{2}{c}{\cfe} \\
\cline{2-3}  \cline{4-5}  \cline{6-7}  \cline{8-9}
\colhead{S/N} & \colhead{$\mu$} & \colhead{$\sigma$} & \colhead{$\mu$} & \colhead{$\sigma$} & \colhead{$\mu$} & \colhead{$\sigma$} & \colhead{$\mu$} & \colhead{$\sigma$}\\
\colhead{} & \colhead{(K)} & \colhead{(K)} & \colhead{(dex)} & \colhead{(dex)} & \colhead{(dex)} & \colhead{(dex)} & \colhead{(dex)} & \colhead{(dex)}}
\startdata
 10.0  &  $+$15 & 172  &  $+$0.086  &  0.680  &  $+$0.327  &  0.415  &  $-$0.162  &  0.485 \\
 12.5  &  $+$18 & 154  &  $+$0.078  &  0.614  &  $+$0.278  &  0.413  &  $-$0.137  &  0.465 \\
 15.0  &  $+$3  & 133  &  $+$0.102  &  0.400  &  $+$0.199  &  0.373  &  $-$0.107  &  0.353 \\
 20.0  &  $-$4  & 117  &  $+$0.033  &  0.335  &  $+$0.124  &  0.352  &  $-$0.074  &  0.355 \\
 25.0  &  $+$5  & 107  &  $+$0.022  &  0.345  &  $+$0.074  &  0.344  &  $-$0.023  &  0.332 \\
 30.0  &  $+$4  & 103  &  $+$0.005  &  0.297  &  $+$0.046  &  0.330  &  $-$0.012  &  0.310 \\
 35.0  &  $+$3  & 104  &  $-$0.004  &  0.259  &  $+$0.035  &  0.325  &  $-$0.004  &  0.324 \\
 40.0  &  $+$10 &  97  &  $+$0.000  &  0.213  &  $+$0.024  &  0.313  &  $+$0.002  &  0.309 
\enddata
\tablecomments{The symbol $\mu$ is the Gaussian mean in the residuals between the SSPP and the high-resolution values, while $\sigma$ 
is calculated following Equations (\ref{eqn:sys}) and (\ref{eqn:tot}). These are derived after adjusting 
for offsets of 52 K for \teff, --0.735 dex for \logg, 0.177 dex for \feh, and 0.128 dex for \cfe, 
found in Figures \ref{fig:hipar} and \ref{fig:hicfe}, respectively.}
\label{tab:noise}
\end{deluxetable}

As far as \logg\ and \feh\ are concerned, the offset generally increases, with larger
scatter, as the quality of the spectrum decreases, again as expected. It
is also seen that the mean offset in \cfe\ becomes larger (in the sense
of an underestimate of [C/Fe]) at low S/N. This presumably arises
because, as a higher level of noise affects the region of the CH $G$-band,
the feature becomes more washed out, resulting in a lower estimate of
\cfe. At S/N = 15 \AA$^{-1}$, it appears that the size of the offsets
for \logg, \feh, and \cfe\ is less than 0.12 dex, which is smaller
than the scatters listed in the table. The scatters for those three
parameters are $\leq 0.4$ dex for S/N $\geq$ 15 \AA$^{-1}$

One useful insight provided by this noise-injection experiment is that,
at high S/N, the dominant error in the total uncertainty is the
external error, $\langle\sigma_{\rm ext}\rangle$ in Equation (\ref{eqn:tot}),
while at low S/N both the external and random error, $\sigma_{\rm
SSPP}$ in Equation (\ref{eqn:tot}) contribute to the total error, as the
size of the scatter become larger with declining S/N, as can 
be seen in Table \ref{tab:noise}.

We conclude from the noise-injection tests performed above that we are
able to estimate \cfe\ with a precision of $\sim$0.35 dex down to S/N =
15 \AA$^{-1}$, while reproducing \feh\ estimates to better than 0.4 dex,
with systematic offsets that are much smaller ($\sim0.15$ dex). Previous
experience suggests that noise experiments of the sort we have carried
out are actually quite conservative in their predictions, so that the
actual scatters in our estimates of [C/Fe] and [Fe/H] are likely to be
smaller than indicated by these experiments.

There is one additional point that we need to address concerning the
above results. We might expect that the error in the determination of
\cfe~would vary with the metallicity of a star, such that the uncertainty of
\cfe~will become larger in more metal-poor than metal-rich stars,
especially for low carbon-abundance levels, as the CH $G$-band decreases
in strength. Another small effect is the increase of atomic blending in
the CH $G$-band region with increasing metallicity, which may also
contribute to the uncertainty of \cfe. Additional noise in the spectrum
of a metal-poor star would drive the uncertainty to even higher values.
As we validate our methods with a sample mostly comprising stars with
[Fe/H] $< -2.5$, the associated errors of \feh\ and \cfe\ might be
larger in the stars with \feh\ $<-3.0$ than with \feh\ $ > -3.0$ under
the same S/N conditions.

In addition, owing to the scarcity of metal-rich stars in our comparison
sample, we are not able to carry out a thorough test on the dependency
of the uncertainty in the measured \cfe~with metallicity. However, we
expect that the error associated with the \cfe\ measurement for the
stars with \feh\ $> -2.5$ will not be larger than the scatter (about 0.3
dex) found in Table \ref{tab:hiressam}, as the metal-rich stars possess
much stronger CH $G$-band features.

\section{The Carbon-Enhanced SDSS/SEGUE Stars}

\subsection{The SDSS/SEGUE Stellar Sample}

All of the SDSS/SEGUE stellar spectra are processed through the latest
version of the SSPP to obtain \teff, \logg, \feh, and \cfe. In order to
assemble a sample with reliable atmospheric parameters and \cfe\ to
analyze the nature of CEMP stars in the field, we first exclude all
stars located on plug-plates that were taken in the direction of known
open and globular clusters. For stars that were observed multiple times
(these are often calibration or quality assurance stars), we retain only
the spectrum with the highest S/N.

\begin{figure*}[t]
\centering
\includegraphics[scale=0.6]{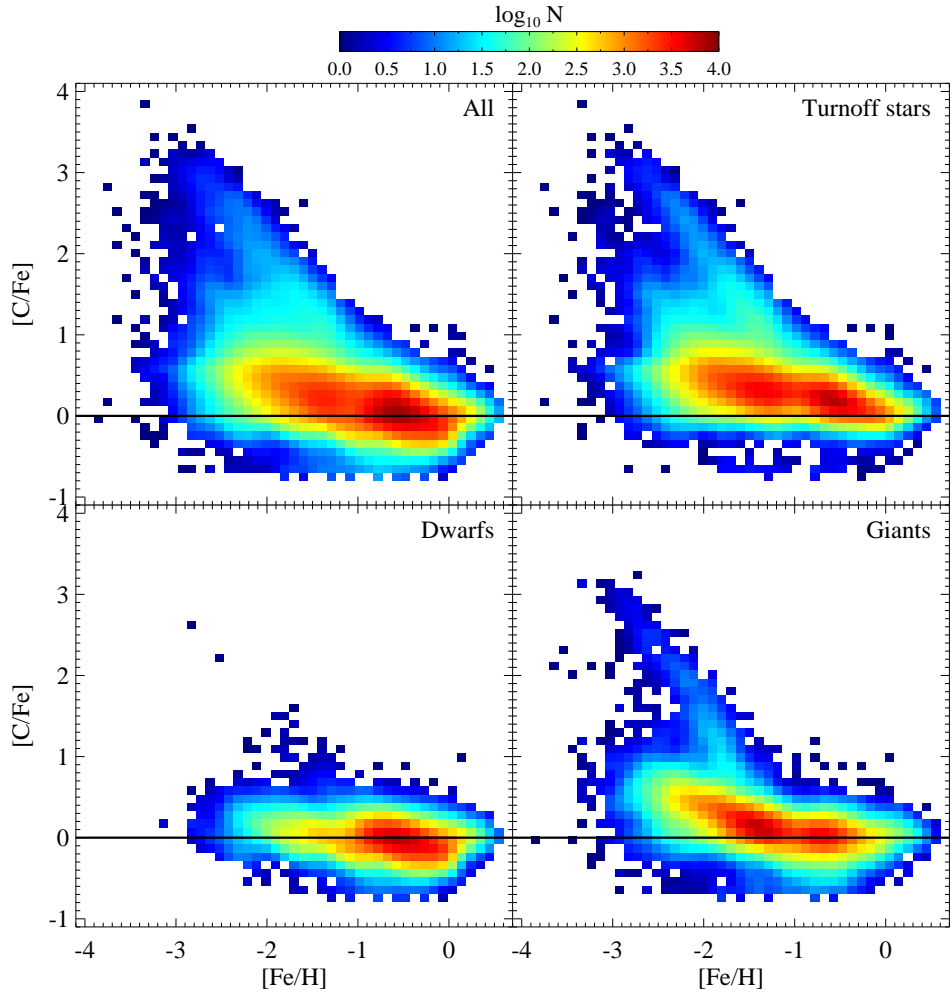}
\caption{Number--density distribution of the SDSS/SEGUE sample 
in the \cfe\ and \feh\ plane, smoothed by a Gaussian kernel, for the
entire sample (top-left panel), the turnoff stars (top-right panel), dwarfs
(bottom-left panel), and giants (bottom-right panel). The turnoff stars are located
in the temperature range 5600~K $\leq$ \teff\ $\leq$ 6700~K. The dwarfs are
occupy the ranges 4400~K $\leq$ \teff\ $<$ 5600~K and \logg\ $\geq 4.0$,
whereas the giant sample corresponds to the regions of 4400~K $\leq$
\teff\ $<$ 5600~K and \logg\ $< 4.0$. The solid horizontal lines are the
solar value of \cfe. The color bar at the top shows the number of stars
per 0.1 $\times$ 0.1 dex bin.}
\label{fig:cempmap}
\end{figure*}

Next, we remove all stars lacking information on their stellar
parameters and \cfe, which can occur for a variety of reasons, but often
because of defects in the spectra. We then apply the following
(conservative) cuts to the sample: S/N $\geq 20$ \AA$^{-1}$, 4400~K
$\leq$ \teff\ $\leq$ 6700~K, and --4.0 $\leq$ \feh\ $\leq +0.5$, so that
our estimate of \cfe\ is as reliable as possible. Finally, we visually
inspect individual spectra with [Fe/H] $\leq -2.0$, to eliminate objects
such as cool white dwarfs or stars with emission-line features in
the cores of their Ca\,{\sc ii} lines, which can produce spurious
low-metallicity estimates from the SSPP. This visual inspection also
removes a small number of additional defective spectra, which sometimes
produces an incorrect metallicity estimate. In addition, we inspect the
spectra for which the SSPP assigns \cfe\ $\geq +0.7$ for [Fe/H] $>-2.0$,
and remove stars with poor estimates of \feh\ and/or \cfe. Furthermore, 
in our analysis of the CEMP stars below, we do not take into account 
the stars with unknown carbon status, which include those with the 
U (upper limit) flag raised and [C/Fe] $\geq +0.7$. There are 
about 1390 such stars. After application of these procedures, 
we end up with a total sample of about 247,350 stars. 

We reiterate that, in our analysis of C-rich stars, we make use of the
SSPP adopted metallicity, [Fe/H]$_{\rm adopted}$ and [C/Fe]$_{\rm
adjusted}$, computed from [C/H] -- [Fe/H]$_{\rm adopted}$, where 
[C/H] = [C/Fe] $+$ [Fe/H] through the carbon-determination routine;
below we simply refer to our final abundance ratios as \cfe\ and \feh. 

The top-left panel of Figure \ref{fig:cempmap} is a logarithmic
number-density map of the full sample of stars with accepted parameter
estimates in the \cfe\ versus \feh\ plane, after smoothing with a Gaussian
kernel; each pixel is 0.1 dex by 0.1 dex. One notable feature seen in
the panel is that, as the metallicity decreases, the distribution of
\cfe\ becomes gradually broader from \feh\ $< -0.5$ and \cfe\ $\sim+0.7$, 
indicating that there exists a greater fraction of C-rich stars among
the metal-poor stars. For this reason, we adopt the C-rich criterion of
\cfe\ $\geq +0.7$ in order for a star to be considered a CEMP star.
Another interesting aspect of the panel is the dramatic increase seen in
the number of stars with \cfe\ $> +2.0$ below \feh\ $ = -2.0$. That is,
at a given metallicity below this value, the distribution of stars has a
longer extended tail to high \cfe.

Additionally, the top-left panel shows a slightly increasing trend of 
\cfe\ for the stars in the ranges \feh\ $< -1.5$ and \cfe\ $< +1.0$. 
In order to investigate what kinds of objects contribute to this
feature, and to be sure that it is not an artifact produced by
incorrect \cfe\ estimation, we divide our sample into main-sequence
turnoff stars (top-right panel) with 5600~K $\leq$ \teff\ $\leq$ 6700~K,
dwarfs (bottom-left panel) with 4400~K $\leq$ \teff\ $<$ 5600~K and \logg\
$\geq$ 4.0, and giants (bottom-right panel) with 4400~K $\leq$ \teff\ $<$
5600~K and \logg\ $<$ 4.0. It is clear that both the turnoff stars and
giant stars contribute to the feature, and it is not produced by
spurious measurement of \cfe\ for some particular type of stars. However,
compared to the dwarfs and giants, the turnoff stars with \feh\ $> -1.0$
and \cfe\ $ < +0.5$ tend to exhibit somewhat higher \cfe, by 0.2--0.3
dex. This level of the offset may be due not only to the uncertainty
of the gravity estimate, owing to the relatively more difficult
determination of \logg\ for such stars, but also to the generally
weaker CH $G$-band features that are found for the warmer turnoff stars.  

\begin{figure*}
\centering
\includegraphics[scale=0.6]{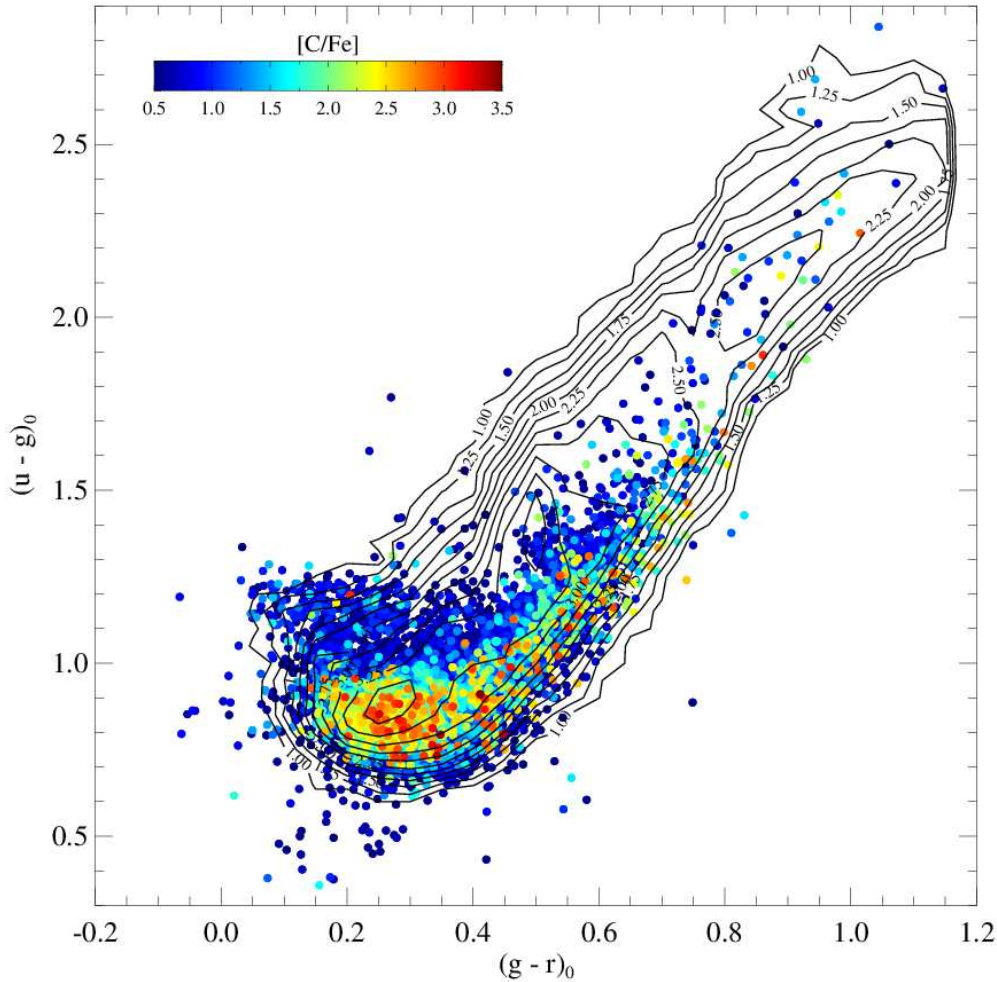}
\caption{Distribution of the stars in the reddening-corrected color--color plane $(u-g)_{0}$ 
and $(g-r)_{0}$ and with \cfe\ $\geq +0.5$. The 
contours delineate the logarithmic number per 0.05
$\times$ 0.05 dex bin for the stars with \cfe\ $< +0.5$. The color bar
provides the scale for the carbon abundance ratios. Because most of the C-rich stars are
metal-poor, the distribution is shifted toward the low side of the
stellar locus. However, as all of the stars with different carbon
enhancements occupy similar regions in the color--color plot, target
selection in SDSS/SEGUE by color neither favors nor disfavors inclusion
of the C-rich stars or the C-normal stars.}
\label{fig:uggr}
\end{figure*}

While the dwarf sample in the bottom-left panel of
Figure~\ref{fig:cempmap} does not exhibit any unusual features, the
bottom-right panel for the giants has a very intriguing branch of
high-[C/Fe] stars below \feh\ $<-2.0$---a very well defined correlation
between [C/Fe] and [Fe/H]. As Masseron et al. (2010) and Spite et al.
(2013) pointed out, this might imply that, regardless of the metallicity
range below \feh\ $<-2.0$, the stars in this branch may possess the same
amount of carbon (similar [C/H]). In other words, there could exist a
limit on the carbon abundances of material transferred from a progenitor
AGB companion, due to the mass range of AGB stars that can produce and
dredge up carbon-enriched material to their surfaces.

Another interesting property from the turnoff sample (top-right panel of
Figure \ref{fig:cempmap}) is that it may be possible to separate the
stars with \feh\ $< -2.5$ into three groups: C-normal with
\cfe\ $< +0.7$, C-intermediate with $+0.7 <$ \cfe\ $< +2.0$, and C-rich
with \cfe\ $> +2.0$. Interestingly enough, this feature does not arise 
for the dwarfs and giants. The intriguing features seen 
among the giants and turnoff stars requires further 
investigation, which is presently underway. 

We have learned from the noise-added synthetic spectra that it is 
difficult to measure the carbon-to-iron ratio for hot, low-gravity stars 
with low carbon abundances, particularly for \teff\ $> 6000$~K and 
\logg\ $< 3.0$. Our full sample of 247,350 objects includes only about
1600 stars with \teff\ $> 6000$~K and \logg\ $< 3.0$, only about
0.6\%. Thus, the impact on our analysis of the CEMP frequency is
minimal.

Before using the SDSS/SEGUE stellar spectra to study the frequency of
the CEMP phenomenon, it is necessary to ensure that the target selection
by colors (e.g., $g-r$ and $u-g$) used in the SDSS/SEGUE does not bias
toward or against the selection of carbon-rich stars, as the strong 
CH $G$-band may influence (primarily) the observed $g$ magnitude. To check
on this possible selection bias, we construct a color-color plot in $(g-r)
_{0}$ and $(u-g)_{0}$, shown in Figure \ref{fig:uggr}. The contours 
delineate the logarithmic number for the stars with \cfe\ $< +0.5$, while 
the filled circles represent the stars with \cfe\ $\geq +0.5$. The 
carbon enhancement is color-coded, and its scale is shown at the top as 
a color bar. Since the C-rich stars are mostly occupied by metal-poor 
stars, the distribution is biased to the lower side of the 
stellar locus in the figure. From inspection of the figure, the 
stars with different carbon enhancements clearly occupy similar regions,
without any isolated loci, in the color-color diagram, suggesting that
the color selection of the targets in the SDSS/SEGUE does not
preferentially select carbon-rich or carbon-normal stars.

\subsection{The Frequency of CEMP Stars as a Function of [Fe/H]}

\subsubsection{Previous Studies}

Spectroscopic follow-up of metal-poor candidates selected from the HK
and HES surveys have identified a number of CEMP stars, and there have
been numerous studies that attempted to derive their frequency, based on
a number of different criteria. Different authors have employed minimum
carbon-abundance ratios for CEMP stars of [C/Fe] $\geq +0.5$, $\geq
+0.7$, or $\geq +1.0$. Aoki et al. (2007) and others have also included
an additional luminosity criterion, in order to account for the
reduction of [C/Fe] in advanced evolutionary stages along the red giant
branch.

Based on a high-resolution spectroscopic analysis of 122 HES metal-poor
giants, Cohen et al. (2005) claimed that the fraction of the CEMP stars
(\cfe\ $\geq +1.0$) is 14.4\%$\pm$4\% for \feh\ $\leq -2.0$. Frebel et al
(2006) also derived a low CEMP fraction (\cfe\ $\geq +1.0$), 9\%$\pm$2\%,
from analysis of medium-resolution spectra for 145 VMP HES giants. 
They also found an increasing trend of CEMP frequency
with distance above the Galactic plane. The claim was clearly confirmed by
Carollo et al. (2012) from an analysis of a much larger sample of
SDSS/SEGUE calibration stars.

On the other hand, based on high-resolution spectroscopy of 349 HES
metal-poor stars from the HERES Survey (Christlieb et al. 2004; Barklem
et al. 2005), Lucatello et al. (2006) calculated a lower limit of
21\%$\pm$2\% for CEMP stars (\cfe\ $\geq +1.0$) for the stars with
\feh\ $\leq -2.0$, higher than the previous two studies. This
discrepancy may result from different sample selections, such as
including a larger fraction of warm stars (making carbon features more
difficult to detect) in the previous studies.

Similar to the present study, Carollo et al. (2012) determined [C/Fe]
employing a $\chi^{2}$ minimization approach using spectral matching
over the CH $G$-band region, and applied their technique to
spectrophotometric and reddening standard stars from SDSS/SEGUE,
resulting in about 31,200 stars with measured carbon-abundance ratios.
Unlike our method, which allows two parameters to vary during the
$\chi^{2}$ minimization step (\feh\ and \cfe), they varied only \cfe. 
Adopting the C-rich definition of [C/Fe] $\geq +0.7$, they obtained a
cumulative frequency of 8\% for [Fe/H] $\leq -1.5$, 12\% for [Fe/H]
$\leq -2.0$, and 20\% for [Fe/H] $\leq -2.5$. They also showed that the
enhancement of carbon relative to iron increases with declining
metallicity, as the average \cfe\ ($\langle$[C/Fe]$\rangle$ $\sim +1.0$) for CEMP
stars at [Fe/H] = --1.5 grows to $\langle$[C/Fe]$\rangle$ $\sim +1.7$ at [Fe/H] =
--2.7.

From an analysis of 25 giants in a sample of stars with [Fe/H] $\leq
-2.5$ among 137 EMP (\feh\ $\leq -3.0$) candidates
selected from SDSS/SEGUE, Aoki et al. (2013) derived a CEMP fraction
(defined using [C/Fe] $\geq +0.7$) as high as 36\%, substantially larger
than any of the previous studies. However, they found only 10 CEMP stars
out of 108 turnoff stars, yielding a fraction of 9\%, which they argue
is a lower limit due to the much weaker features of the CH $G$-band in
these warmer stars.

Yong et al. (2013) have reported, for a large sample of halo
stars with available high-resolution spectroscopy, that the C-rich
population represents 32\%$\pm$8\% of stars below [Fe/H] = $-3.0$,
(again using [C/Fe] $\geq +0.7$ as their CEMP criterion). These previous
studies allow us to infer that the fraction of CEMP stars discovered to
date roughly rises from 10\% to 20\% for \feh\ $<-2.0$ to 30\% for
[Fe/H] $< -3.0$, 40\% for [Fe/H] $< -3.5$, and 75\% for [Fe/H] $< -4.5$
(Beers \& Christlieb 2005; Norris et al. 2007, 2013), after inclusion of
a recently recognized star with \feh\ $\sim -5.0$ and \cfe\ $\leq +0.7$
(which is not carbon enhanced, at least according to the [C/Fe] $\geq
+1.0$ criterion; see Caffau et al. 2011).

\subsubsection{Present Results}

In order to investigate the previously noted trends with metallicity in
more detail, we now derive the frequencies of CEMP stars based on the 
large sample of SDSS/SEGUE spectra. We note that, owing to the low
resolution of the SDSS/SEGUE spectra, the accuracy of our determination
of \cfe\ is less than that based on high-resolution analysis. However,
as our sample size is so much larger than previous studies, we expect to
produce meaningful new results for the trends in CEMP frequency.

For clearly detected C-enhanced stars, we only count as C-rich objects
those stars with \cfe\ $\geq +0.7$, a correlation coefficient (CC) at
least 0.7, and lacking an upper limit flag (`U'). This sample of stars
is regarded as $N_{\rm C}$. The CC is calculated by comparing the
observed and synthetic spectrum over 4290--4318 \AA. Then, the
cumulative frequency of C-enhanced objects ($F_{\rm C}$) is computed by
dividing the number of C-rich stars ($N_{\rm C}$) by all stars ($N_{\rm
total}$), counted below a given metallicity ([Fe/H] = $+0.0, -0.5, -1.0,
-1.5, -2.0, -2.5, -3.0, -3.5$). In the form of an equation, 

\begin{equation}
{F_{\rm C}} = \frac{N_{\rm C}([\rm C/\rm Fe] \geq +0.7, \rm CC \geq 0.7, D~or~L~flag)}{N_{\rm total}}.
\label{eqn:frac}
\end{equation}

\noindent Note that $N_{\rm total}$ in the denominator of the 
above expression includes all stars in the numerator, plus stars with
[C/Fe] $\geq +0.7$ and CC $ < 0.7$, indicating a poor carbon
measurement, the C-normal stars (D or L flag, [C/Fe] $< +0.7$),
independent of the value of CC, and the stars with a U flag raised
having [C/Fe] $< +0.7$, again independent of the value of CC. Stars with
unknown carbon status, which include those with the U flag raised and
[C/Fe] $\geq +0.7$, are not included in the above definition.

\begin{deluxetable*}{crcrcrcr|rcrcrcr}
\centering
\tablewidth{0pc}
\setlength{\tabcolsep}{0.001in}
\tabletypesize{\scriptsize}
\tablecaption{Cumulative Frequencies of CEMP
Stars for Three Different Carbon Abundance Criteria}
\tablehead{\colhead{} & \multicolumn{7}{c}{SDSS/SEGUE $+$ Literature Sample\tablenotemark{1}} & \multicolumn{7}{c}{SDSS/SEGUE} \\
\cline{2-8}  \cline{9-15} 
\colhead{} & \multicolumn{2}{c}{[C/Fe]$\ge+0.5$}  &
\multicolumn{2}{c}{[C/Fe]$\ge+0.7$}  & \multicolumn{2}{c}{[C/Fe]$\ge+1.0$} & \colhead{} &
\multicolumn{2}{c}{[C/Fe]$\ge+0.5$}  &
\multicolumn{2}{c}{[C/Fe]$\ge+0.7$}  & \multicolumn{2}{c}{[C/Fe]$\ge+1.0$} & \colhead{} \\
\colhead{[Fe/H]} & \colhead{$N_{\rm C}$} &  \colhead{$F_{\rm C}$}  & \colhead{$N_{\rm C}$} &  \colhead{$F_{\rm C}$}  & \colhead{$N_{\rm C}$} &  \colhead{$F_{\rm C}$}  & \colhead{$N_{\rm total}$} &
\colhead{$N_{\rm C}$} &  \colhead{$F_{\rm C}$}  & \colhead{$N_{\rm C}$} &  \colhead{$F_{\rm C}$}  & \colhead{$N_{\rm C}$} &  \colhead{$F_{\rm C}$}  & \colhead{$N_{\rm total}$}}
\startdata

 $\leq  +0.0$ &  11622&  0.05$\pm$0.01 &   5001&  0.02$\pm$0.01 &   2641&  0.01$\pm$0.01 & 243653&  11581&  0.05$\pm$0.01 &   4967&  0.02$\pm$0.01 &   2612&  0.01$\pm$0.01 & 243577\\
 $\leq  -0.5$ &  11548&  0.06$\pm$0.01 &   4996&  0.03$\pm$0.01 &   2640&  0.01$\pm$0.01 & 191252&  11507&  0.06$\pm$0.01 &   4962&  0.03$\pm$0.01 &   2611&  0.01$\pm$0.01 & 191176\\
 $\leq  -1.0$ &  10792&  0.11$\pm$0.01 &   4861&  0.05$\pm$0.01 &   2627&  0.03$\pm$0.01 &  96974&  10751&  0.11$\pm$0.01 &   4827&  0.05$\pm$0.01 &   2598&  0.03$\pm$0.01 &  96898\\
 $\leq  -1.5$ &   8386&  0.16$\pm$0.01 &   3914&  0.08$\pm$0.01 &   2222&  0.04$\pm$0.01 &  51107&   8345&  0.16$\pm$0.01 &   3880&  0.08$\pm$0.01 &   2193&  0.04$\pm$0.01 &  51031\\
 $\leq  -2.0$ &   3799&  0.25$\pm$0.01 &   2029&  0.13$\pm$0.01 &   1171&  0.08$\pm$0.01 &  15500&   3758&  0.24$\pm$0.01 &   1995&  0.13$\pm$0.01 &   1142&  0.07$\pm$0.01 &  15424\\
 $\leq  -2.5$ &    775&  0.30$\pm$0.01 &    549&  0.21$\pm$0.01 &    378&  0.15$\pm$0.01 &   2587&    734&  0.29$\pm$0.01 &    515&  0.21$\pm$0.01 &    349&  0.14$\pm$0.01 &   2511\\
 $\leq  -3.0$ &    106&  0.34$\pm$0.03 &     89&  0.28$\pm$0.03 &     70&  0.22$\pm$0.03 &    314&     65&  0.27$\pm$0.03 &     55&  0.23$\pm$0.03 &     41&  0.17$\pm$0.03 &    238\\
 $\leq  -3.5$ &     21&  0.57$\pm$0.12 &     16&  0.43$\pm$0.11 &     15&  0.41$\pm$0.10 &     37&      2&  0.25$\pm$0.18 &      2&  0.25$\pm$0.18 &      2&  0.25$\pm$0.18 &      8
\enddata
\tablecomments{$N_{\rm C}$ is the number of stars within each
metallicity range and with \cfe\ $\geq$ $+$0.5, $+$0.7, or $+$1.0, and 
$F_{\rm C}$ is the fraction of the stars with \cfe\ $\geq$ $+$0.5, $+$0.7, or $+$1.0, 
calculated from $N_{\rm C}/N_{\rm total}$, see text. The quoted error is
derived from Poisson statistics. If the fraction of CEMP stars and its associated error 
is less than 1\%, we assume to have at least 1\%.}
\tablenotetext{1}{The additional literature values mostly come from Table 1 of Yong et al. (2013), 
and all of these are extremely metal-poor stars ([Fe/H] $< -3.0$). 
In their table, we remove four SDSS stars and we adopt the parameters
from a dwarf-star (rather than subgiant) analysis for eight of their stars. One object from Caffau et al.
(2011) is also included, and this star is removed from the SDSS/SEGUE sample.}
\label{tab:cemp}
\end{deluxetable*}

This approach to estimation of the frequency of CEMP stars is essentially
the same as in Equation (2) of Carollo et al. (2012), except that they
used the CH $G$-band strength ($>1.2$ \AA) to indicate a clear detection
of the CH $G$-band, rather than the CC criterion used in
this study. As in Carollo et al. (2012), our calculated CEMP fraction is
a lower limit, since some likely bona-fide CEMP stars with CC $< 0.7$
are undercounted in the numerator of Equation (3) above, due to a
poor spectrum (or with a poor match to the synthetic spectrum), and
appear in the denominator instead.

\begin{figure}
\centering
\plotone{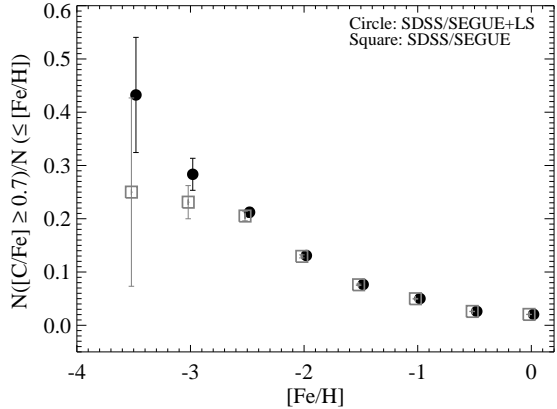}
\caption{Cumulative frequencies of CEMP stars (\cfe\ $\geq +0.7$) in
different metallicity ranges. [Fe/H] indicates the upper limit of the
metallicity range considered at each plotted point. The open squares are the
frequencies obtained by considering only the SDSS/SEGUE sample, while
the filled circles include the literature sample (LS) from Yong et al.
(2013) and Caffau et al. (2011). Poisson error bars are plotted. For
clarity, the $X$-axis (\feh) values of
the open squares are shifted by $-0.02$ dex, and the filled circles are
shifted by $+0.02$ dex.}
\label{fig:cemp}
\end{figure}

Figure \ref{fig:cemp} shows the derived cumulative frequency of CEMP
stars versus \feh, with their associated Poisson error bars. The open
squares indicate the calculation when considering only our SDSS/SEGUE
stars, while the filled circles represent the fraction when including 
stars from the LS (mostly, Table 1 of Yong et al.
2013 and one object from Caffau et al. 2011), based on high-resolution
analyses. Note that, in order to improve visibility, the $X$-axis (\feh)
values of the open squares are shifted by $-0.02$ dex, while the filled
circles are shifted by $+0.02$ dex.

The motivation for including the LS stars is to increase the number of
stars with \feh\ $ < -3.0$, for better number statistics in the
derivation of the CEMP frequency in the lowest-metallicity regime.
Before using the Yong et al. stars, we first remove the four SDSS/SEGUE
stars in their sample. Some of the Yong et al. stars have two sets of
\feh\ and \cfe\ estimates reported; one from a dwarf analysis and the
other from a subgiant analysis. We adopt the \feh\ and \cfe\ values from
the analysis under the assumption of dwarf luminosity classification for
those stars to increase the number of dwarf stars. There are seven such
stars, and they all have \cfe\ $> +1.0$. No large differences in \cfe\
and \feh\ exist between the dwarf and subgiant analyses.

Note also that, as the object from Caffau et al. (2011) in the LS
is also one of the SDSS/SEGUE stars, we remove it from our SDSS/SEGUE sample
to avoid double counting of this object.\footnote[18]{The reason for
excluding the star from our SDSS/SEGUE sample, but not from the LS, is
that the SSPP-estimated [Fe/H] of --3.79 is too high compared to [Fe/H]
$\sim -5.0$ by Caffau et al. (2011).} In Figure \ref{fig:cemp}, the
Poisson error bars are sufficiently large to be visible only for [Fe/H]
$\leq -3.0$. Table \ref{tab:cemp} lists all derived quantities,
including results obtained when adopting the CEMP criteria of \cfe\
$\geq +0.5$ and $\geq +1.0$, for completeness. In the table, if the
fraction of the CEMP stars or its associated error is less than 1\%, we
assume it to be 1\%.

For \feh\ $ <-3.0$, the sample including the LS stars exhibits a higher
fraction of CEMP stars than the SDSS/SEGUE-only sample, although the
error bars between the two samples overlap. This implies there are more
C-rich stars ([C/Fe] $\geq +0.7$) for [Fe/H] $<-3.0$ in the LS than 
in the SDSS/SEGUE sample. Note also that, as listed in Table
\ref{tab:cemp}, there are only eight stars (two of which are CEMP stars) with
[Fe/H] $\leq -3.5$ available, which precludes derivation of a
statistically meaningful frequency for the SDSS/SEGUE sample at this
metallicity.

When only considering the high-resolution sample from the
literature, Yong et al. (2013) obtained a CEMP frequency of 32\%$\pm$8\%
for stars with \feh\ $\leq -3.0$ and \cfe\ $\geq +0.7$, which is larger by 
9\% than our derived value from the SDSS/SEGUE sample alone
(23\%$\pm$3\%), but marginally compatible with theirs to within the
error bars. When adopting \cfe\ $\geq +1.0$ as the criterion for a CEMP star,
Yong et al. (2013) estimated a frequency of 23\%$\pm$6\%, while our
value is 17\%$\pm$3\%, which agrees well within the error bars. 

Inspection of Figure \ref{fig:cemp} reveals that the cumulative CEMP
frequency for the SDSS/SEGUE$+$LS sample (circle symbols in the figure)
rises slowly from 2\% at [Fe/H] $\leq 0.0$ to about 13\% at [Fe/H] $\leq
-2.0$ (close to that reported by Carollo et al. 2012), followed by a
more rapid increase from [Fe/H] $\leq -2.0$ to [Fe/H] $\leq -3.5$. This
trend does not differ when adopting other definitions for carbon
enhancement (\cfe\ $\geq +0.5$ or $\geq +1.0$), as seen in Table
\ref{tab:cemp}. It appears that our derived value of the CEMP frequency
(8\%$\pm$1\% for \feh\ $\leq -2.0$ and \cfe\ $\geq +1.0$) is closer
to those of Cohen et al. (2005) and Frebel et al. (2006) than to that of
Lucatello et al. (2006), but again, one must recall the possible
selection effect in making this comparison. Our result for the
cumulative frequency of CEMP stars ([C/Fe] $\geq +0.7$) with [Fe/H]
$\leq -2.5$, 21\%$\pm$1\%, matches well with that obtained by Carollo et
al. (20\%).

Yong et al. (2013) examined the possibility that the CEMP fraction
as a function of \feh\ continues to rise with declining metallicity below
[Fe/H ] = --3.0. After dividing their objects with $-4.5 \leq$ \feh\
$\leq -3.0$ into three bins with similar numbers of stars, they
calculated the CEMP fractions in those three bins, and derived a slope
of --0.24$\pm$0.22 for the CEMP frequencies for the stars with \feh\
$\leq -3.0$, which included three stars with \feh\ $< -4.5$. The inclusion
of the star (with \feh\ $\sim -5.0$ with \cfe\ $\leq +0.7$) from Caffau et
al. (2011) yields a slope of --0.20$\pm$0.19. Based on these results,
they concluded that there was no significant correlation between the
fraction of CEMP stars and \feh\ among stars of the lowest metallicity.

\begin{figure}
\centering
\plotone{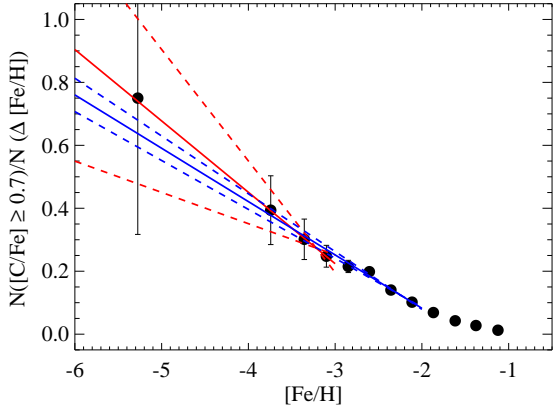}
\caption{Differential frequencies of CEMP stars (\cfe\ $\geq +0.7$) as a
function of \feh. The bin with \feh\ $< -5.0$ has only four stars, as
listed in Table \ref{tab:cempslope}, one of which is from Caffau et al.
(2011), and assumed to be a C-normal star. The red-solid
line is the slope obtained by fitting the frequencies for \feh\ $\leq
-3.0$, while the blue-solid line is the slope derived from the
sub-sample in the range \feh\ $ < -2.0$. Poisson 
error bars are plotted. The dashed lines indicate the 1$\sigma$ 
errors in the derived slopes after taking into account the measured Poisson 
error in the calculated frequencies.}
\label{fig:cempslope}
\end{figure}

Figure \ref{fig:cempslope} shows the differential frequencies of CEMP
stars in each bin of \feh\ from the SDSS/SEGUE$+$LS sample. Table
\ref{tab:cempslope} lists the metallicity bins used, average metallicity
in each bin, and the fraction of the CEMP stars in each metallicity bin.
The observed trend of the differential CEMP frequency from the figure is that it
steadily increases from 1\% at \feh\ $\sim-1.0$ to 75\% at \feh\
$\sim-5.25$.

Calculating the slopes of the CEMP fractions from these observations, we
obtain a slope of --0.23$\pm0.13$ from a linear fit to the fractions
with \feh\ $\leq -3.0$, shown as a red-solid line in the figure. The
dashed lines are 1$\sigma$ errors in the computed slope (note that the
measured Poisson error in the fraction of each metallicity bin is taken
into account during the fit). This value is in good agreement from that of
Yong et al. (--0.20, after the inclusion of Caffau et al.'s star).
Extending the metallicity range up to \feh\ = --2.0, the derived slope
is --0.17$\pm0.01$, shown as a blue-solid line in the figure, which is
not far from that of the more metal-poor region, and certainly is
consistent to within the allowed errors. Thus, this overall behavior
suggests there exists at most a mildly increasing differential frequency
of CEMP stars with decreasing metallicity. However, because there are
not many stars below \feh\ = --3.5 discovered to date, more objects are
required to obtain confident estimates of the CEMP frequency at the
lowest metallicities.

\begin{deluxetable}{ccrrc}
\tabletypesize{\scriptsize}
\centering
\tablewidth{0pc}
\setlength{\tabcolsep}{0.001in}
\tablecaption{Differential Frequencies of
CEMP Stars in Bins of Metallicity from the SDSS/SEGUE and Literature\tablenotemark{1} Samples}
\tablehead{\colhead{\feh\ Range} & \colhead{$\langle$[Fe/H]$\rangle$} & \colhead{$N_{\rm C}$} &  \colhead{$N_{\rm total}$} & \colhead{$F_{\rm C}$}}
\startdata

 $-1.25 \leq [\rm Fe/H] < -1.00$& $-1.12$ &    262&  20876&   0.01$\pm$0.01\\
 $-1.50 \leq [\rm Fe/H] < -1.25$& $-1.38$ &    685&  24991&   0.03$\pm$0.01\\
 $-1.75 \leq [\rm Fe/H] < -1.50$& $-1.62$ &    907&  21359&   0.04$\pm$0.01\\
 $-2.00 \leq [\rm Fe/H] < -1.75$& $-1.87$ &    978&  14248&   0.07$\pm$0.01\\
 $-2.25 \leq [\rm Fe/H] < -2.00$& $-2.11$ &    873&   8584&   0.10$\pm$0.01\\
 $-2.50 \leq [\rm Fe/H] < -2.25$& $-2.36$ &    607&   4329&   0.14$\pm$0.01\\
 $-2.75 \leq [\rm Fe/H] < -2.50$& $-2.61$ &    339&   1708&   0.20$\pm$0.01\\
 $-3.00 \leq [\rm Fe/H] < -2.75$& $-2.85$ &    122&    567&   0.22$\pm$0.02\\
 $-3.25 \leq [\rm Fe/H] < -3.00$& $-3.10$ &     51&    206&   0.25$\pm$0.03\\
 $-3.50 \leq [\rm Fe/H] < -3.25$& $-3.36$ &     22&     73&   0.30$\pm$0.06\\
 $-4.50 \leq [\rm Fe/H] < -3.50$& $-3.74$ &     13&     33&   0.39$\pm$0.11\\
 $-6.00 \leq [\rm Fe/H] < -4.50$& $-5.28$ &  3\tablenotemark{2}&      4&   0.75$\pm$0.43
\enddata
\tablecomments{$\langle$[Fe/H]$\rangle$ is an average of \feh\ in each metallicity range. $N_{\rm C}$ is the number of C-rich 
stars (\cfe\ $\geq 0.7$) within each metallicity range, and 
$F_{\rm C}$ is the frequency of the CEMP stars calculated by $N_{\rm
C}/N_{\rm total}$, see text. The quoted error is derived from Poisson statistics. 
If the fraction of the CEMP stars and its associated error is less than
1\%, we assume it to be at least 1\%.}
\tablenotetext{1}{The added literature values are the same as in Table \ref{tab:cemp}.}
\tablenotetext{2}{The object from Caffau et al. (2011) is assumed to be a carbon-normal star.}
\label{tab:cempslope}
\end{deluxetable}

Nonetheless, Figure \ref{fig:cempslope} provides us with an overall
accurate trend of steadily increasing CEMP fractions with decreasing
metallicity, confirming and extending the results of previous studies,
with much improved Poisson errors compared to past efforts. Another
important point evident from inspection of the figure is that the
differential frequency of CEMP stars may change somewhat rapidly from
one metallicity bin to another, especially for [Fe/H] $< -2.0$, hence as
``fine-grained'' a sample as possible is required to be sensitive to
this behavior, which may contain clues to the nature of the progenitor
stellar populations of CEMP stars.
 
\subsection{Frequencies of CEMP Stars among Giants, Turnoff Stars, and
Dwarfs} 

Aoki et al. (2013) derived a CEMP frequency (defined using [C/Fe] $\geq
+0.7$) as high as 36\% for their giant stars, but a frequency of only
9\% was obtained for their main-sequence turnoff stars (which they
considered a lower limit). One caution to be aware of in their analysis
is that there might be a bias in their high-resolution spectroscopic
sample, since they made use of the CH $G$-band to select CEMP candidates,
although it is much weaker for the warmer turnoff stars. Thus, their
frequency estimate for stars near the turnoff will only include stars
with higher carbon abundance ratios (e.g., \cfe\ $\sim +1.5$ for \feh\
$\sim -3.0$; see Aoki et al. 2013). 

On the other hand, Yong et al. (2013) also obtained a higher CEMP
fraction for dwarfs in their sample (50\%$\pm$31\%) than for giants
(39\%$\pm$11\%), although these estimates overlap within the errors. These
results suggest the possibility that the fraction of the CEMP stars
varies between different luminosity classes. 

We now examine how the cumulative frequency of CEMP stars differs among
giants, turnoff stars, and dwarfs in our SDSS/SEGUE+LS sample. The
classifications used for each category are---giant: 1.0 $<$ \logg\
$\leq$ 3.5, turnoff: 3.5 $<$ \logg\ $\leq$ 4.2, and dwarf: 4.2 $<$
\logg\ $\leq$ 5.0. We stress again that we adopt the parameters from the
dwarf analysis in the LS from Yong et al. (2013) in order
to increase the size of the dwarf sample. Had we taken the assumed gravity 
for the subgiant analysis, six additional stars would have been included in the
turnoff sub-sample, while one object would belong to the giant
sub-sample, instead of all being dwarfs in our luminosity
classifications; the CEMP frequencies would change accordingly.

Because the turnoff and giant stars probe more distant regions of the
Galaxy than the dwarfs, the possibility exists that the change of the
CEMP frequency may be influenced by changes in the mix of populations
(inner halo versus outer halo). Thus, we also investigate the CEMP fraction
for stars located within 5 kpc of the Galactic mid-plane (i.e., $|Z| <
5$ kpc). This distance condition rejects some of the giant and turnoff
stars, but includes most of the dwarfs.

\begin{figure}
\centering
\plotone{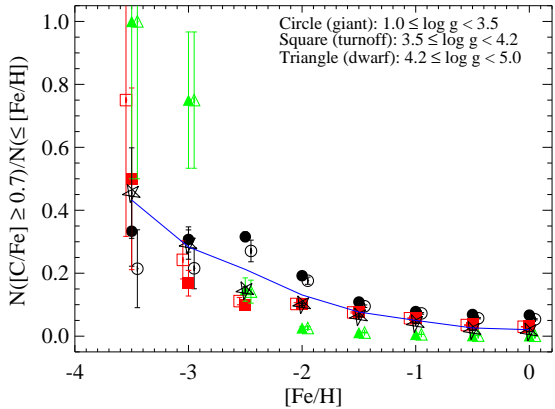}
\caption{Cumulative frequencies of CEMP stars (\cfe\ $\geq +0.7$)
from the SDSS/SEGUE$+$LS sample, as a function of [Fe/H], for three
different luminosity classes: giants (circles), main-sequence turnoff
stars (squares), and dwarfs (triangles). We assume classifications based
on the measured surface gravity---giant: $1.0 \leq$ \logg\ $< 3.5$,
turnoff: $3.5 \leq$ \logg\ $< 4.2$, and dwarf: $4.2 \leq$ \logg\ $<
5.0$. Poisson error bars are plotted. For an alternative comparison, we also
restrict the sample to stars within 5 kpc from the Galactic mid-plane; the open
symbols represent the frequencies from the distance-restricted sample.
The star symbols are the frequency of CEMP stars from all stars with
$|Z| < 5$ kpc. The blue-solid line shows the frequencies calculated in
Figure \ref{fig:cemp}, without any restriction on distance from the
plane. For clarity, the $X$-axis (\feh) values of the open triangles and circles are shifted
by $+0.05$ dex, while the open squares are shifted by $-0.05$ dex.}
\label{fig:cempgra}
\end{figure}

Figure \ref{fig:cempgra} shows the changes in the cumulative fractions
of CEMP stars with \feh. The circles represent giants, the squares the
turnoff stars, and the triangles the dwarfs. The open symbols denote the
sample with $|Z| < 5$ kpc. The star symbols indicate the CEMP
frequencies, without the luminosity classifications, within $|Z|$ = 5
kpc. We follow Equation \ref{eqn:frac} to compute the fraction for each
population. The computed fraction for each sub-sample over various
metallicity ranges is listed in Table \ref{tab:cempgra} for the entire
sample, and for the sub-sample restricted by the vertical distance
criterion. Note that, for clarity, the $X$-axis (\feh) values of the open
triangles and circles are shifted by $+0.05$ dex, whereas the open
squares are shifted by $-0.05$ dex.

Examining first the filled symbols in Figure \ref{fig:cempgra}, which are
derived without application of the distance cut, the general 
trend of the cumulative frequency of the CEMP giants is quite
intriguing.  Unlike the steady increase of the CEMP frequency for 
the turnoff stars and dwarfs, the CEMP fraction of the giants
does {\it not increase}, but remains almost at the same value below
[Fe/H] = -- 2.5. This may be due to the expected dilution of carbon by
CNO-processed material from the interiors of these stars, as is argued
to have occurred for other samples of metal-poor giants by Spite et al.
(2006).

The cumulative frequency of CEMP giants for \feh\ $\leq -2.5$ (32\%$\pm$2\%)
is in good agreement with that of Aoki et al. (2013) (36\%$\pm$12\%),
while the fraction for \feh\ $\leq\ -3.0$ (31\%$\pm$4\%) is slightly
lower than that of Yong et al. (2013) for their giant sample
(39\%$\pm$11\%). However, all of these measurements are close enough to
be consistent within the claimed Poisson errors. The cumulative CEMP 
frequency (12\%$\pm$1\%) of our giant sample in the
range of \feh\ $\leq -2.0$ (and using \cfe\ $\geq +1.0$) is closer to that 
reported by Cohen et al. (2005), 14\%$\pm$4\%, than that by Frebel 
et al. (2006), 9\%$\pm$2\%.

For the main-sequence turnoff stars, the figure indicates that the 
CEMP frequency initially slowly increases with decreasing metallicity,
but exhibits a rapid increase below [Fe/H] = --2.5. We obtain a frequency of
10\%$\pm$1\% for \feh\ $\leq -2.5$, which is in good agreement with that
reported by Aoki at al. (2013), 9\%$\pm$3\%. 

The observed change in the cumulative frequency of CEMP stars for the
dwarfs is rather dramatic. The dwarf population exhibits a very rapid
increase in the fraction of the CEMP stars below [Fe/H] = --2.5, jumping
from about 15\%$\pm$4\% at \feh\ $\leq -2.5$ to 75\%$\pm$22\% at \feh\
$\leq -3.0$. Our sample suggests that all dwarfs are CEMP stars below
\feh\ = --3.5, but this is based on a sample of only four stars. In
fact, these all come from the previous studies compiled by Yong et al.
(2013). A larger sample of dwarfs in this metallicity range will be
needed in order to confirm this behavior.

We now consider the effect that application of a distance restriction
($|Z| < 5$ kpc) to the sample has on the derived CEMP frequencies. For
giant and turnoff stars, the CEMP fraction changes rather significantly
for \feh\ $\leq -2.5$. Inspection of Figure \ref{fig:cempgra} reveals
that the frequency of the CEMP giants within 5 kpc of the Galactic
mid-plane (open circles) does not appear to increase below \feh\ =
--2.5, but rather, levels off to a roughly constant value. By
comparison, the CEMP frequencies of the turnoff sample within $|Z| <
5$ kpc (open squares) becomes even higher than that of the entire
turnoff sample (filled squares) below [Fe/H] = $-3.0$. This may suggest
that the mix of a greater fraction of EMP (\feh\
$<-3.0$)  turnoff stars relative to giants present among the stars with $|Z| < 5$
kpc would result in an overall higher frequency of CEMP stars. 
This effect can be confirmed from the observed cumulative CEMP fractions,
computed from all stars within $|Z| = 5$ kpc.

\tabletypesize{\tiny}
\begin{deluxetable*}{crrc|rrc|rrc|rrc|rrc|rrc}
\tablewidth{0pc}
\renewcommand{\tabcolsep}{0in} \tablecaption{Cumulative Frequencies of
CEMP Stars Classified as Giants, Turnoff Stars, and Dwarfs from the SDSS/SEGUE and Literature\tablenotemark{1} Samples}
\tablehead{\colhead{} & \multicolumn{9}{c}{All} & \multicolumn{9}{c}{$|Z| < 5$ kpc} \\
\hline
\colhead{} & \multicolumn{3}{c}{Giant} & \multicolumn{3}{c}{Turnoff}  & \multicolumn{3}{c}{Dwarf} & \multicolumn{3}{c}{Giant} & \multicolumn{3}{c}{Turnoff}  & \multicolumn{3}{c}{Dwarf} \\
\cline{2-4}  \cline{5-7} \cline{8-10} \cline{11-13}  \cline{14-16} \cline{17-19}
\colhead{[Fe/H]} & \colhead{$N_{\rm C}$} & \colhead{$N_{\rm total}$} & \colhead{$F_{\rm C}$}  & \colhead{$N_{\rm C}$} & \colhead{$N_{\rm total}$} & 
\colhead{$F_{\rm C}$}  & \colhead{$N_{\rm C}$} & \colhead{$N_{\rm total}$} & \colhead{$F_{\rm C}$} & \colhead{$N_{\rm C}$} & \colhead{$N_{\rm total}$} & \colhead{$F_{\rm C}$}  & \colhead{$N_{\rm C}$} & \colhead{$N_{\rm total}$} & 
\colhead{$F_{\rm C}$}  & \colhead{$N_{\rm C}$} & \colhead{$N_{\rm total}$} & \colhead{$F_{\rm C}$}}
\startdata
$\leq  +0.0$ &2206&32995&0.07$\pm$0.01 &2629&80898&0.03$\pm$0.01 &159&129648&0.01$\pm$0.01 & 793&14802&0.05$\pm$0.01 &2242&74910&0.03$\pm$0.01 &156&128959&0.01$\pm$0.01 \\
$\leq  -0.5$ &2205&32042&0.07$\pm$0.01 &2627&69327&0.04$\pm$0.01 &157& 89772&0.01$\pm$0.01 & 792&13916&0.06$\pm$0.01 &2240&63515&0.04$\pm$0.01 &154& 89216&0.01$\pm$0.01 \\
$\leq  -1.0$ &2174&27817&0.08$\pm$0.01 &2530&42905&0.06$\pm$0.01 &150& 26141&0.01$\pm$0.01 & 763&10543&0.07$\pm$0.01 &2150&37750&0.06$\pm$0.01 &147& 25879&0.01$\pm$0.01 \\
$\leq  -1.5$ &1916&17652&0.11$\pm$0.01 &1886&24244&0.08$\pm$0.01 &105&  9111&0.01$\pm$0.01 & 554& 5827&0.10$\pm$0.01 &1586&21044&0.08$\pm$0.01 &102&  9031&0.01$\pm$0.01 \\
$\leq  -2.0$ &1179& 6138&0.19$\pm$0.01 & 804& 7804&0.10$\pm$0.01 & 39&  1484&0.03$\pm$0.01 & 239& 1357&0.18$\pm$0.01 & 656& 6414&0.10$\pm$0.01 & 38&  1476&0.03$\pm$0.01 \\
$\leq  -2.5$ & 420& 1330&0.32$\pm$0.02 & 110& 1117&0.10$\pm$0.01 & 16&   108&0.15$\pm$0.04 &  62&  229&0.27$\pm$0.03 &  90&  808&0.11$\pm$0.01 & 15&   106&0.14$\pm$0.04 \\
$\leq  -3.0$ &  59&  192&0.31$\pm$0.04 &  17&  101&0.17$\pm$0.04 & 12&    16&0.75$\pm$0.22 &  11&   51&0.22$\pm$0.07 &  16&   66&0.24$\pm$0.06 & 12&    16&0.75$\pm$0.22 \\
$\leq  -3.5$ &   9&   27&0.33$\pm$0.11 &   3\tablenotemark{2}&6&0.50$\pm$0.29 &  4\tablenotemark{2}&4\tablenotemark{2}&1.00$\pm$0.50 &   3\tablenotemark{2}&14&0.21$\pm$0.12 &3\tablenotemark{2}&    4&0.75$\pm$0.43 &  4\tablenotemark{2}&     4\tablenotemark{2}&1.00$\pm$0.50
\enddata
\tablecomments{Two samples of stars are considered. One includes all stars, while the other one consists of stars within $\pm5$ kpc from the Galactic plane (e.g., $|Z| < 5$ kpc). 
$N_{\rm C}$ is the number of CEMP stars (\cfe\ $\geq 0.7$) within each metallicity range, and $F_{\rm C}$ is the
fraction of the CEMP stars calculated by $N_{\rm C}/N_{\rm total}$, see
text. The quoted error is derived from Poisson statistics. If the
fraction of the CEMP stars and its 
associated error is less than 1\%, we assume it to be at least 1\%. We
consider the stars in the surface-gravity range 1.0 $\leq$ \logg\ $<$ 3.5, 3.5 $\leq$ \logg\
$<$ 4.2, and 4.2 $\leq$ \logg\ $<$ 5.0, to be a giants, turnoff stars, and
dwarfs, respectively. }
\tablenotetext{1}{The added literature values are the same as in Table \ref{tab:cemp}.}
\tablenotetext{2}{No SDSS/SEGUE stars are included in this bin.}
\label{tab:cempgra}
\end{deluxetable*}

Another interesting point to be drawn from Figure \ref{fig:cempgra} is
that, in the metallicity regime $-2.5 \le$ [Fe/H] $\le -1.5$, the CEMP
frequency derived from all stars with $|Z| < 5$ kpc is slightly lower
than the CEMP fraction without application of a distance restriction
(blue-solid line; the same sample shown in Figure \ref{fig:cemp}). This
behavior is consistent with previous demonstrations that the CEMP
frequency increases with increasing distance from the Galactic plane,
indicating that there may exist a greater fraction of CEMP stars
associated with the outer-halo population than the inner-halo
population (Carollo et al. 2012). 

Figure \ref{fig:cempgra} also implies that, as the giants are
intrinsically brighter than the turnoff and dwarf stars (hence are more
likely to be observed beyond $|Z| = 5$ kpc), one might expect a bias
toward lower derived frequencies of CEMP stars if the sample under
consideration extends to include larger volumes of the Galaxy. We
conclude that a volume-limited sample that populates different stellar
luminosity classes as equally as possible should be used to obtain more
meaningful comparisons of their respective CEMP frequencies as a
function of [Fe/H] or $|Z|$. A more detailed analysis and
interpretation of the dependence of the frequency of CEMP stars on their
kinematic properties and spatial distribution will be presented in a
future paper.

\section{Summary and Conclusions}

We have presented a method for estimating \cfe~from the low-resolution
($R = 2000$) SDSS/SEGUE stellar spectra, based on spectral matching
against a custom grid of synthetic spectra. In order to validate our
method, we have performed a star-by-star comparison between the
SDSS/SEGUE spectra of stars with available high-resolution
determinations of \cfe, carried out tests of the impact of
\teff\ and \logg\ errors on the determination of \feh\ and \cfe, and 
conducted noise-injection experiments on both the
synthetic spectra and SDSS/SEGUE stars with literature values of
\cfe\ based on high-resolution spectroscopy, 

Checks on possible errors in our determination of \cfe~due to our
preference of fixing \teff\ and \logg\ at the values delivered by the
SSPP reveals that the mean offsets associated with different input
shifts in \teff\ and \logg\ are mostly smaller than the derived rms
scatter in the determination of \cfe. We confirm that the
surface-gravity error in the SSPP has only a minor impact on our
measured \cfe. Within \teff\ shifts of $\pm$200~K (which is equivalent
to the typical error of the SSPP-determined \teff), our determined \cfe\
is perturbed by less than $\pm$0.25 dex, which is smaller than the 
rms scatter of 0.30 dex, the typical error of our measured \cfe.

Our noise-injection experiments suggests that our approach is capable of
estimating \cfe~with a precision $<$ 0.35 dex for spectra with S/N $\geq
15$ \AA$^{-1}$ over the parameter space \teff~= [4400, 6700] K, \logg~=
[1.0, 5.0], \feh~= [--4.0, $+$0.0], and \cfe~= [$-$0.25, $+3.5$].
According to our noise-injection experiments, errors in the
determination of \cfe~increase to $\sim$0.35 dex for S/N $< 15$
\AA$^{-1}$. Thus, it is recommended to use the spectra with a minimum
S/N = 15 \AA$^{-1}$ for the application of this approach.

The method presented here can be easily applied to other spectra that
cover similar wavelength ranges at similar resolving power. Therefore,
it should be a useful new tool for the investigation of the
chemical-enrichment history of Galactic populations, with stellar
spectra obtained by other large spectroscopic surveys such as LAMOST.

Using the SDSS/SEGUE and LS, we have investigated how the
differential frequency of CEMP stars changes as a function of [Fe/H]. We
find that the CEMP frequency slowly rises from almost zero (1\%) to
about 14\%$\pm$1\% at [Fe/H] $\sim$ --2.4, and there is a marked
increase, by about a factor of three (39\%), from [Fe/H] $\sim$ --2.4
to $\sim$ --3.7. The gradient of the CEMP fraction does not change much 
over different metallicity regimes, suggesting a steady increase of 
the frequency of CEMP stars with decreasing metallicity. However, owing to the 
handful of stars with \feh\ $<$--3.5 identified to date, it is necessary 
to collect more stars in this range to robustly characterize the CEMP 
frequency for extremely and ultra-metal poor stars.

We have also investigated how the cumulative frequency of CEMP stars
varies between different luminosity classes. Unlike the dwarfs and
turnoff stars, which show continuously rising trends below [Fe/H] =
--2.5, the giant sample exhibits a roughly constant CEMP fraction
below [Fe/H] $\leq -2.5$. The giant sample exhibits a fraction of CEMP
stars for \feh\ $\leq -2.5$ of 32\%, which is in good agreement with that
reported by Aoki et al. (2013) (36\%), while the fraction of CEMP giant
stars for \feh\ $\leq -3.0$ (31\%) is somewhat lower than that of Yong
et al. (2013) (39\%). In both cases the Poisson error bars overlap. For
the main-sequence turnoff stars we obtain a CEMP fraction of 10\% for
\feh\ $\leq -2.5$, in excellent agreement with that of Aoki at al.
(2013) (9\%). However, as Aoki et al. (2013) illustrated, there remains
the difficulty of identifying CEMP stars in this temperature regime
due to the low resolution of the SDSS/SEGUE spectra, such that the CH
$G$-band can only be detected for stars with higher carbon abundances.
Lastly, the dwarf stars exhibit a very rapid increase in the cumulative
frequency of CEMP stars below [Fe/H] = --2.5, leaping from a fraction of
15\% at \feh\ $\leq -2.5$ to about 75\% at \feh\ $\leq$ --3.0. All of
the dwarfs with \feh\ $\leq -3.5$ come from previous high-resolution
studies, and all are CEMP stars. Since it is only based on a sample of
four stars, the sample size must be substantially increased in order to
confirm this result. 

Analysis of a distance-restricted sample ($|Z| < 5$ kpc) reveals that
the cumulative frequency of CEMP stars classified as giants {\it does
not appear} to increase with declining metallicity, but rather, remains
roughly constant below [Fe/H] = --2.5. On the other hand, the cumulative
frequencies of the turnoff sample of CEMP stars increases below \feh\ =
--3.0, indicating that a sample biased to include more EMP turnoff stars
than giants (as might arise from examination of a local volume), may
result in an overall trend of higher CEMP frequencies with decreasing
metallicity. As discussed by Lucatello et al. (2006), Spite et al.
(2006), and Aoki et al. (2007), the apparent lack of an increase in the
frequency of CEMP stars among giants could well be associated with extra
mixing of CNO-processed material from their interiors, diluting the
C-rich material in their envelopes. Confirmation of this effect could
come from observations of the $^{12}$C/$^{13}$C and [N/Fe] ratios for such
stars. Inspection of the distance-restricted sample also indirectly
confirms the increasing trend of CEMP frequency with increasing distance
from the Galactic mid-plane, previously pointed out by Frebel et al.
(2006) and Carollo et al. (2012). Future analysis of this sample, taking
into account a more detailed examination of the kinematics and spatial
distribution of these stars, should prove illuminating.   

\acknowledgments
Funding for SDSS-III has been provided by the Alfred P. Sloan Foundation, the 
Participating Institutions, the National Science Foundation, and the U.S. 
Department of Energy Office of Science. The SDSS-III Web site is 
http://www.sdss3.org/.

SDSS-III is managed by the Astrophysical Research Consortium for the 
Participating Institutions of the SDSS-III Collaboration including 
the University of Arizona, the Brazilian Participation Group, 
Brookhaven National Laboratory, University of Cambridge, Carnegie 
Mellon University, University of Florida, the French Participation 
Group, the German Participation Group, Harvard University, the 
Instituto de Astrofisica de Canarias, the Michigan State/Notre 
Dame/JINA Participation Group, Johns Hopkins University, Lawrence 
Berkeley National Laboratory, Max Planck Institute for Astrophysics, 
Max Planck Institute for Extraterrestrial Physics, New Mexico State 
University, New York University, Ohio State University, Pennsylvania 
State University, University of Portsmouth, Princeton University, 
the Spanish Participation Group, University of Tokyo, University of 
Utah, Vanderbilt University, University of Virginia, University of 
Washington, and Yale University. 

Y.S.L. is a Tombaugh Fellow. This work was supported in part by grants
PHY 02-16783 and PHY 08-22648: Physics Frontiers Center/Joint Institute
for Nuclear Astrophysics (JINA), awarded by the U.S. National Science
Foundation. T.M. is supported in part by an Action de Recherche
Concert\'ee from the Direction g\'en\'erale de l'enseignement non
obligatoire et de la Recherche Scientifique, Direction de la Recherche
Scientifique, Communaut\'e Fran\c{c}aise de Belgique and by the
F.R.S.-FNRS FRFC grant 2.4533.09. V.M.P. acknowledges support for this
work through FAPESP fellowship (2012/13722-1).

\end{document}